\documentclass[aps,prb,showpacs,twocolumn,floats,epsfig,pdflatex, superscriptaddress,longbibliography]{revtex4-2}
\usepackage{amssymb}
\usepackage{amsmath}
\usepackage{amsfonts}
\usepackage{graphicx}
\usepackage{amssymb}
\usepackage{float}
\usepackage{hyperref}

\begin{document}

\title{Inherited Berry curvature of phonons in Dirac materials with time-reversal symmetry}
\author{Sayandip Ghosh}
\email{sayandipghosh@phy.vnit.ac.in}
\affiliation{Department of Physics, Visvesvaraya National Institute of Technology Nagpur, Nagpur 440010, India}
\author{Sel\c{c}uk Parlak}
\affiliation{D\'epartement de physique, Institut quantique and Regroupement Qu\'eb\'ecois sur les Mat\'eriaux de Pointe,  Universit\'{e} de Sherbrooke, Sherbrooke, Qu\'{e}bec J1K 2R1, Canada}
\author{Ion Garate}
\email{ion.garate@usherbrooke.ca}
\affiliation{D\'epartement de physique, Institut quantique and Regroupement Qu\'eb\'ecois sur les Mat\'eriaux de Pointe,  Universit\'{e} de Sherbrooke, Sherbrooke, Qu\'{e}bec J1K 2R1, Canada}

\date{\today}

\begin{abstract}
The Berry curvature of phonons is an active subject of research in condensed matter physics. 
Here, we present a model in which phonons acquire a Berry curvature through their coupling to electrons in crystals with time-reversal symmetry.
We illustrate this effect for BaMnSb$_2$, a quasi two-dimensional Dirac insulator, whose low-energy massive Dirac fermions generate a phonon Berry curvature that is proportional to the electronic valley Chern number. We estimate the contribution of the phonon Berry curvature to the thermal Hall conductivity in BaMnSb$_2$.

\end{abstract}

\maketitle

\section{Introduction}
The Berry curvature of phonons in crystals has been introduced and extensively studied in the past decade, giving birth to the field of topological phononics \cite{zhang2010topological, qin2012berry, liu2018berry, liu2020topological, ding2024topological, zhang2018double, miao1018observation}. 
Identifying and classifying materials with nontrivial phononic band geometry is a topic of ongoing interest \cite{jin2018recipe, coh2023classification, xu2024catalog}. One item that has received little attention thus far is the
interplay between the phonon Berry curvature and the Berry curvatures of other quasiparticles of the crystal, notably electrons. 
Nevertheless, understanding this interplay could provide insights about the microscopic origin of the phonon Berry curvature.

From symmetry point of view, the phonon Berry curvature and the electronic Berry curvature are allowed in generic points of the Brillouin zone under the same circumstances, i.e. when space inversion (${\cal P}$) or time-reversal (${\cal T}$) symmetries are broken in the crystal; the product  ${\cal P T}$ must likewise be broken.

From microscopic point of view, however, there is no evident connection between the two curvatures. For instance, it is well-known that the electronic Berry curvature can be nonzero in the absence of phonons \cite{xiao2010berry}. Conversely, models for topological phonons have been conceived, e.g. in two-dimensional honeycomb lattices \cite{liu2017model}, without any reference to electrons. Also, it has been reported that the phonon Berry curvature can be induced by a magnetic field in materials that have no electronic Berry curvature, like Si \cite{saito2019berry}.  But, could there be more complex systems in which phonons would acquire a Berry curvature due solely to their coupling to topologically nontrivial electrons, or vice versa? This type of question has not been widely explored.

Recently, Saparov {\em et al.} \cite{saparov2022lattice} have partially touched on the interplay between the phonon Berry curvature and the electronic band topology by investigating the electron-phonon interaction in the Haldane model. 
They have found that the molecular Berry curvature, which is an electronic Berry curvature resulting from ionic displacements, is proportional to the electronic Chern number (i.e. to the flux of the electronic Berry curvature through the Brillouin zone). This molecular Berry curvature modifies the equation of motion for lattice vibrations, ultimately resulting in a phonon Berry curvature that emerges from the electronic Chern number. 

Crucially, the mechanism for the generation of the phonon Berry curvature in Refs. \cite{saito2019berry} and \cite{saparov2022lattice} relies on  breaking ${\cal T}$ either by magnetic order or by a magnetic field.
Therefore, the question remains whether there could be a link between the phonon Berry curvature and the electronic Berry curvature in crystals that preserve ${\cal T}$. 
In the present study, we answer this question affirmatively. 

To do so, we consider BaMnSb$_2$, a prototype quasi-two-dimensional Dirac insulator that breaks ${\cal P}$ while preserving ${\cal T}$ \cite{liu2021spin}.
The dynamical matrix of this crystal, if described by a phenomenological force-constant model with two-body interactions, leads to phonons (which we call ``bare'' phonons)  with a vanishing Berry curvature.

Upon inclusion of the coupling between lattice vibrations and the low-energy massive Dirac fermions of BaMnSb$_2$, a hybridization takes place between two linearly polarized, orthogonal bare phonon modes.
This hybridization produces ``dressed''  phonons that  are elliptically polarized  \cite{hu2021phonon}, with an angular momentum that is proportional to the electronic valley Chern number (i.e. to the difference between two valley-projected Chern numbers) \footnote{Along a loosely related line of inquiry, Ref. \cite{lin2024quantum} has shown that the quantum oscillations in acoustic phonons are altered by the electronic Berry phase.}.
Because the direction of the angular momentum is locked to the wave vector, the dressed phonons have been characterized as “helical.”
The helical texture of the phonon angular momentum in the Brillouin zone respects ${\cal T}$ and crystal symmetries. As of now, there has been no experimental observation of this phonon helicity, though it could be accessible in the presence of temperature gradients \cite{hamada2018phonon,zhang2024observation} or in Raman spectroscopy \cite{parlak2023detection}.

The Berry curvature of phonons was not considered in Ref. \cite{hu2021phonon}. In the present work, we prove that the helical phonons in BaMnSb$_2$ possess a Berry curvature that is proportional to the electron-phonon coupling and to the electronic valley Chern number. 
Since the bare phonons had no Berry curvature, it is as though phonons ``inherited'' a Berry curvature from their coupling to topologically nontrivial electrons.
This result is qualitatively different from that of Ref. \cite{saparov2022lattice} in that BaMnSb$_2$ preserves ${\cal T}$ and the molecular Berry curvature is zero. 
The finding of an ``inherited'' phonon Berry curvature in topologically nontrivial electron systems with ${\cal T}$ symmetry is the main result from our work.

The rest of this work is organized as follows. 
In Secs. \ref{sec:prelim} and \ref{sec:bare_phonons} we review the key electronic and lattice vibrational properties of BaMnSb$_2$.
One insight from Sec.  \ref{sec:bare_phonons} is that bare, long-wavelength optical and acoustic phonons have a vanishing Berry curvature because the force constant model is symmetric under the exchange of cartesian coordinates. 
In Sec. \ref{sec:e-ph}, we consider the coupling between long-wavelength phonons and massive Dirac fermions.
The effect of that coupling in the dynamical matrix and in the phonon Berry curvature is studied in Sec. \ref{sec:dressed_phonons}. 
It is found that the microscopic mechanism that produces phonon helicity  is also responsible for the phonon Berry curvature.
In Sec. \ref{sec:TR_breaking}, we consider the influence of a ${\cal T}$-breaking perturbation on the phonon Berry curvature. 
In a magnetic field, the ``inherited'' phonon Berry curvature is no longer an odd function of the phonon wave vector and therefore contributes to the thermal Hall effect, whose value we estimate. 
Some technical details are collected in the appendices. Concerning units, $\hbar\equiv 1$ is adopted throughout the text, with the exception of Sec. \ref{sec:the}.

\section{Preliminaries}
\label{sec:prelim}
We begin by briefly reviewing relevant aspects of BaMnSb$_2$, which we take here as a prototype Dirac material with time-reversal symmetry. It is a layered material with a quasi-two-dimensional (quasi-2D) electronic structure with low-energy bands being very weakly dispersive along $(001)$ direction \cite{liu2021spin}. 
The $xy$ crystal planes in which the Sb atoms are located form 2D rectangular lattices with two Sb atoms (Sb$_1$ and Sb$_2$) per unit cell. Distortion shifts Sb$_2$ atoms away from the center of the rectangles formed by Sb$_1$ atoms, resulting in a series of zigzag chains.  This distortion stabilizes the lattice structure~\cite{chen2024thermoelectric} and results in the point group  $C_{2v}$, spanned by a two-fold rotation $C_{2x}$ along the $x$-axis, a mirror plane $\sigma_v(xz)$ perpendicular to the $y$ axis, and a mirror plane $\sigma_v(xy)$ perpendicular to the $z$ axis. 

The electronic band structure near the Fermi energy consists of two massive Dirac cones, related to one another by time-reversal symmetry, and situated at momenta (valleys) $K_+$ and $K_-$.
These bands primarily come from the $p_x$ and $p_y$ orbitals of Sb atoms.
The effective Hamiltonian \cite{hu2021phonon}  around a valley $s$ is
\begin{equation}
{\cal H}^s_{\rm el} = \sum_{\bf k} c^\dagger_{s, \alpha, {\bf k} } h^s_{\alpha\beta} ({\bf k})  c_{s, \beta, {\bf k} },
\end{equation}
where $\alpha,\beta \in\{p_x, p_y\}$ are orbital labels, $c$ and $c^\dagger$ are the electronic annihilation and creation operators, and $h^s_{\alpha\beta}$ is the matrix element of the ${\bf k}\cdot{\bf p}$ Hamiltonian
\begin{equation}
\label{eq:hek}
h^s ({\bf k}) = s \left[v_0 (k_x \tau_x + k_y \tau_z) + m_0 \tau_y \right].
\end{equation}
Here, $v_0$ is the Dirac velocity, $m_0$ is the Dirac mass induced by the zig-zag distortion, and $\tau_{0,x,y,z}$ are Pauli matrices for the orbital index.  
Within this low-energy subspace, $\sigma_v (xy)$ corresponds to an identity matrix and does not
change the 2D momentum. Thus, we only consider $C_{2x}$ and time-reversal  (${\cal T}$) operations henceforth. The Hamiltonian transforms as $C_{2x} h^s_e (k_x, k_y)C_{2x}^{-1} = h^{-s}_e (k_x, - k_y )$ and ${\cal T} h^s ({\bf k}){\cal T}^{-1} = h^{-s} (-{\bf k})$, where the symmetry representations are $C_{2x} = \tau_z$ and ${\cal T}= \tau_0 {\cal K}$ (${\cal K}$ stands for complex conjugation). Within one valley, we only have the combined symmetry $C_{2x} {\cal T} = \tau_z {\cal K}$.

\section{Bare in-plane phonons}
\label{sec:bare_phonons}

In this section we calculate the Berry curvature of lattice vibrations involving the in-plane displacements of Sb ions. 
By ``bare'', we mean phonons that are not coupled to low-energy Dirac fermions. 
We model those phonons via phenomenological short-range interatomic force constants that are symmetric under the interchange of cartesian coordinates.
The coupling to Dirac fermions will be incorporated in Secs. \ref{sec:e-ph} and \ref{sec:dressed_phonons}, upon which the phonons will be called ``dressed''.

\subsection{Dynamical matrix}
\begin{figure}[t]
\begin{center}
\includegraphics[width=\columnwidth]{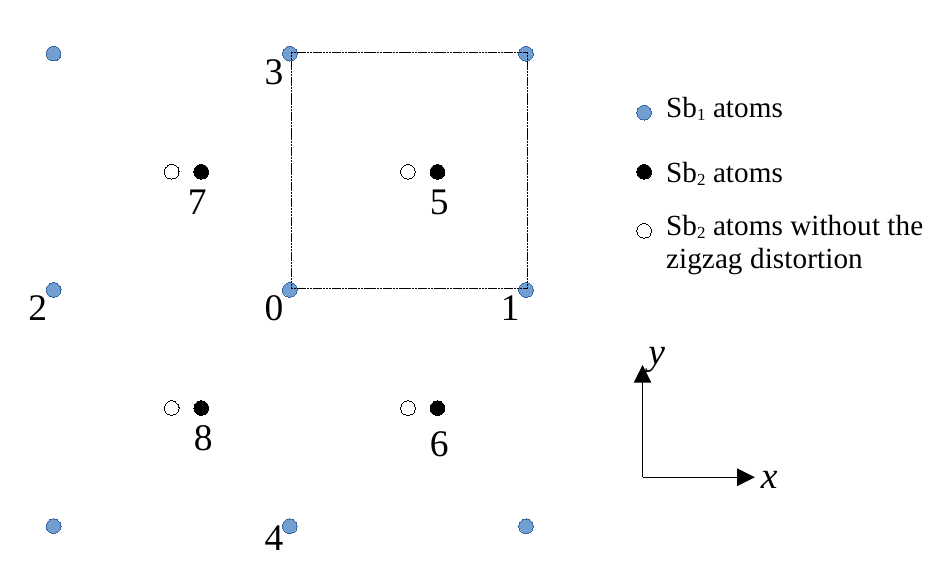}
\caption{Layer of Sb atoms in BaMnSb$_2$. The rectangular unit cell, of dimensions $a\times b$, is shown by the dotted lines. For simplicity, we take $a=b$ in the main text. This simplification does not result in qualitative changes to our main results.}
\label{fig:BaMnSb2}
\end{center}
\end{figure}

The crystal structure of a Sb layer in the $xy$ plane is illustrated in Fig.~\ref{fig:BaMnSb2}, with $\delta a/2$ the atom displacement due to the zig-zag distortion of the lattice (the distance between the white and the black dots, in units of the lattice constant $a$, so that $\delta a$ is dimensionless). 
Owing to the two Sb atoms in the 2D unit cell, there exist six phonon modes (three acoustic and three optical) associated with vibrations of Sb atoms. 
These modes are characterized by the displacements of the two atoms away from their equilibrium positions, $(u_{1x},u_{1y},u_{1z},u_{2x},u_{2y},u_{2z})$.
In this work, we disregard the phonon modes with vibrations along the $z$-axis, since they are odd under mirror reflection $\sigma_v(xy)$ and therefore cannot couple to electronic states near Fermi level, which are even under $\sigma_v(xy)$.

Following Ref. \cite{hu2021phonon}, the bare in-plane lattice vibrations are {\em defined} through phenomenological force constants 
 \begin{align}
 \label{eq:centralpotential}
\Phi_{\alpha\beta}(l \kappa,l^\prime \kappa^\prime)=\gamma(l\kappa,l^\prime \kappa^\prime)\hat{\bf e}_{\alpha}\hat{\bf e}_{\beta},
\end{align}
where $(l \kappa)$ labels an atom $\kappa$ in unit cell $l$, $\alpha,\beta\in\{x,y\}$, $\hat{\bf e}$ is the unit vector in the direction of the vector connecting  atoms $(l \kappa)$ and $(l' \kappa')$,  and $\gamma(l\kappa,l^\prime\kappa^\prime)$ is the spring constant between atoms $(l \kappa)$ and $(l' \kappa')$.
Using the atom labels of Fig.~\ref{fig:BaMnSb2}, the spring constants involving atom $0$ are named as
\begin{align}
\gamma_0 &\text{ (between atom $0$ and atoms $1,2,3,4$)}\notag\\ 
\gamma_1  &\text{ (between atom $0$ and atoms $7,8$)}\notag\\
\gamma_2  &\text{ (between  atom $0$ and atoms $5,6$)}.
\end{align}
The spring constants involving an atom different from $0$ can be constructed similarly. 
The couplings between farther neighbors are neglected.

We have also tried more general models for the bare force constants (cf. App. \ref{ap:dynamical}), verifying that they do not bring about qualitative changes to the main results of our paper. One key property of these phenomenological models is that 
\begin{equation}
\label{eq:force_constant_symmetry}
\Phi_{\alpha\beta}(l \kappa,l^\prime \kappa^\prime)=\Phi_{\beta\alpha}(l \kappa,l^\prime \kappa^\prime),
\end{equation}
 which, although known to hold for two-body forces \cite{michel2011phonon}, is more restrictive than the always-true condition  $\Phi_{\alpha\beta}(l \kappa,l^\prime \kappa^\prime)=\Phi_{\beta\alpha}(l' \kappa',l \kappa)$ \cite{maradudin1968symmetry}.

The bare phonon polarizations and frequencies are obtained from the eigenvectors and the (square-roots of) eigenvalues of the dynamical matrix, respectively. The elements of the dynamical matrix are given by
\begin{equation}
\label{eq:dyn_mat}
D_{\alpha \beta}(\kappa,\kappa^\prime;{\bf q})=\frac{\sum_{l^\prime} \Phi_{\alpha \beta}(l \kappa ,l^\prime \kappa^\prime) e^{-i {\bf q}.{\bf x}_{l\kappa,l'\kappa'}}}{\sqrt{M_{\kappa}M_{\kappa^\prime}}},
\end{equation}
where  ${\bf x}_{l\kappa,l'\kappa'}={\bf x}(l \kappa)-{\bf x}(l^\prime \kappa^\prime)$, ${\bf x}(l\kappa)={\bf x}(l)+{\bf x}(\kappa)$ is the position of atom $(l\kappa)$, ${\bf x}(l)$ is the position of unit cell $l$ and ${\bf x}(\kappa)$ is the position of atom $\kappa$ with respect to the origin in unit cell $l$, $M_{\kappa}$ is the mass of atom $\kappa$ and ${\bf q}=(q_x, q_y)$ is the phonon wave vector. 

The definition in Eq.~(\ref{eq:dyn_mat}) differs from that of Ref. \cite{maradudin1968symmetry} in the phase factor: we use ${\bf x}(l \kappa)-{\bf x}(l^\prime \kappa^\prime)$ in the exponent, instead of ${\bf x}(l)-{\bf x}(l')$. 
This difference in convention is physically inconsequential, but does affect some technical steps below.
For example, the dynamical matrix in our case is not periodic under a translation by a reciprocal lattice vector ${\bf G}$:
\begin{align}
  \label{eq:nonper}
  D_{\alpha \beta}(\kappa,\kappa^\prime; {\bf q}+{\bf G})&=e^{-i{\bf G}\cdot({\bf x}(\kappa)-{\bf x}(\kappa^\prime))} D_{\alpha \beta}(\kappa,\kappa^\prime; {\bf q}).
\end{align}

From the elements $D_{\alpha \beta}(\kappa,\kappa^\prime; {\bf q})$, the dynamical matrix in the $(u_{1x},  u_{1y},  u_{2x} , u_{2y})$ basis is constructed as
\begin{equation}
\label{eq:dyn}
D({\bf q}) = \begin{pmatrix}
D_{xx}(1,1) & D_{xy}(1,1) & D_{xx}(1,2) & D_{xy}(1,2) \\
 & D_{yy}(1,1) & D_{yx}(1,2) & D_{yy}(1,2)\\
 & & D_{xx}(2,2) & D_{xy}(2,2)\\
& & & D_{yy}(2,2)
\end{pmatrix},
\end{equation}
where the lower-triangular terms in Eq. (\ref{eq:dyn}) are determined from the hermiticity condition $D({\bf q})=D^\dagger({\bf q})$. 
The ${\bf q}$-dependence of the matrix elements in Eq.~(\ref{eq:dyn}) has been left implicit for brevity of notation.
The explicit expressions for $D_{\alpha \beta}(\kappa,\kappa^\prime;{\bf q})$ were derived in Ref.  \cite{hu2021phonon} from Eq.~(\ref{eq:centralpotential})  and are collected in App. \ref{ap:dynamical} for reference. 
 
 \begin{figure}[t]
\begin{center}
\includegraphics[width=\columnwidth]{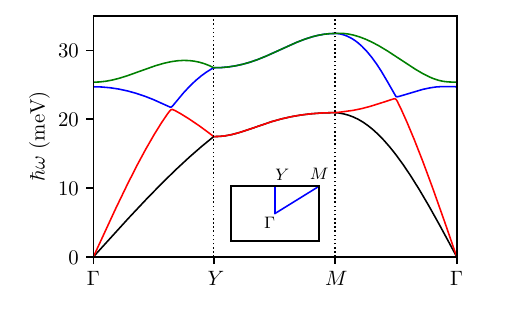}
\caption{Phonon dispersion along high symmetry directions indicated with blue triangle in the inset. The numerical values of the parameters are $\gamma_0=50$ N/m, $\gamma_1=75$ N/m, $\gamma_2=70$ N/m, $M_1=M_2=2 \times 10^{-25}$ kg, $\delta a=0.2$. The scale of phonon frequencies is chosen to qualitatively agree with first-principle calculations~\cite{chen2024thermoelectric}.}
\label{fig:phonondispersion}
\end{center}
\end{figure}

Phonon bands calculated from Eq.~(\ref{eq:dyn}) are shown in Fig. \ref{fig:phonondispersion}. One salient feature in the plot is the double degeneracy of the phonon dispersion along the YM line of the Brillouin zone edge (cf. Fig.~\ref{fig:phonondispersion}), which, as we explain next, can be understood from the symmetry properties of the dynamical matrix. 
 
 Under a symmetry operation $R$, the dynamical matrix transforms as
 \begin{align}
\label{eq:vosko}
\boldsymbol{\Gamma}({\bf q};R) D({\bf q}) \boldsymbol{\Gamma}^{-1}({\bf q}; R)  = D({\bf R}\cdot{\bf q}),
\end{align}
 where ${\bf R}$ is the (proper or improper) rotation matrix associated to the symmetry operation $R$ and $\boldsymbol{\Gamma}$ is the representation matrix for the symmetry operator $R$. 
An explicit expression for $\boldsymbol{\Gamma}$ is given in Eq. (2.37) of Ref. \cite{maradudin1968symmetry}, whose phase factor we must modify  due to the different convention used in Eq.~(\ref{eq:dyn_mat}).
Repeating the procedure of Ref. \cite{maradudin1968symmetry} with our convention, we find the phase factor of $\boldsymbol{\Gamma}$ to be $i({\bf R}\cdot {\bf q})\cdot ({\bf v}+{\bf x}(m))$, where ${\bf v}$ is the fractional translation associated to the symmetry operation $R$ (${\bf v}=0$ for a symmorphic operation) and ${\bf x}(m)$ is a translation vector of the crystal.
 
In case of the $C_{2x}$ and $\sigma_v (xz)$ symmetries of BaMnSb$_2$, we have 
\begin{equation}
\label{eq:gamma_C2x}
\boldsymbol{\Gamma}({\bf q};C_{2x}) = \begin{pmatrix}
1 & 0& 0& 0\\
0& - 1 & 0&0\\
0&0 & 1&0\\
0&0 &0 & - 1
\end{pmatrix}
\end{equation} 
in the $(u_{1x}, u_{1y}, u_{2x}, u_{2y})$ basis, as well as ${\bf R}\cdot{\bf q} = (q_x, -q_y)$.
With this, it is easy to verify that the dynamical matrix elements shown in App. \ref{ap:dynamical} satisfy Eq.~(\ref{eq:vosko}).
But this alone cannot explain the degeneracy of the phonon bands along YM.
It turns out that the degeneracy is a consequence of an additional glide plane symmetry $G_{2y}$ present in the lattice of Fig.~\ref{fig:BaMnSb2} (despite it being absent in the space group of the 3D BaMnSb$_2$ crystal).

The operation $G_{2y}$ consists of a translation along $y$ by half of a unit cell, followed by a mirror reflection perpendicular to the $x$ axis.
The symmetry exchanges the two Sb atoms in the two dimensional unit cell. Then, 
\begin{equation}
\label{eq:gamma_G2y}
   \boldsymbol{\Gamma}({\bf q};G_{2y}) =e^{-iq_y b/2}\begin{pmatrix}
0 & 0 & -1 & 0\\
0 & 0 & 0 & 1\\
-1 & 0 & 0 & 0\\
0 & 1 & 0 & 0
\end{pmatrix},
\end{equation}
 where $b$ is the lattice constant in the $y$ direction. Also, ${\bf R}\cdot{\bf q} = (-q_x, q_y)$.

The last important symmetry to understand the degeneracy of the phonon modes along the YM line is ${\cal T}$-symmetry, which implies \cite{maradudin1968symmetry}
\begin{equation}
\label{eq:Tsym}
D({\bf q})=D(-{\bf q})^*.
\end{equation}

Combining Eqs.~(\ref{eq:nonper}),  (\ref{eq:vosko}) and (\ref{eq:gamma_C2x}), the dynamical matrix along the YM line acquires the form
 \begin{align}
\label{eq:DYMC2x}
\begin{pmatrix}
   D_{xx}(1,1) & 0& 0& D_{xy}(1,2)  \\
   0 & D_{yy}(1,1)  &  D_{yx}(1,2) & 0\\
   0 & D_{yx}(1,2)^* & D_{xx}(2,2)&0 \\
 D_{xy}(1,2)^*& 0 & 0 &D_{yy}(2,2) \\
\end{pmatrix},
\end{align}
which can be immediately block diagonalized into two $2\times 2$ blocks.
We will now prove that the two blocks are identical because $D_{\alpha\alpha}(1,1; {\bf q})=D_{\alpha\alpha}(2,2;  {\bf q})$ and $D_{\alpha\beta}(1,2; {\bf q})=D_{\beta\alpha}(1,2;  {\bf q})$, thereby explaining the two double-degenerate phonon bands at the YM line in Fig.~\ref{fig:phonondispersion}.
To show this, we notice the following symmetry relations:
\begin{align}
\label{eq:diag}
  &D_{\alpha \alpha}(\kappa,\kappa; q_x, q_y) \overset{G_{2y}}= D_{ \alpha \alpha}(\bar{\kappa},\bar{\kappa}; -q_x, q_y) \nonumber \\ 
    &\overset{\cal T} =D_{ \alpha \alpha}(\bar{\kappa},\bar{\kappa}; q_x, -q_y)^* \overset{C_{2x}}= D_{ \alpha \alpha}(\bar{\kappa},\bar{\kappa}; q_x, q_y)^* \nonumber\\
    &\overset{\text{hermiticity}}= D_{ \alpha \alpha}(\bar{\kappa},\bar{\kappa}; q_x, q_y),
\end{align}
where $\kappa$ and $\bar{\kappa}$ are different atoms (i.e. $\kappa=1$ if $\bar{\kappa}=2$, and $\kappa=2$ if $\bar{\kappa}=1$).
The first equality of Eq.~(\ref{eq:diag}) has been derived from combining Eqs. (\ref{eq:vosko}) and (\ref{eq:gamma_G2y}).
Thus,  Eq.~(\ref{eq:diag})  proves that $D_{\alpha\alpha}(1,1; {\bf q})=D_{\alpha\alpha}(2,2;  {\bf q})$.  
Similarly, $D_{\alpha\beta}(1,2; {\bf q})=D_{\beta\alpha}(1,2;  {\bf q})$ follows directly from Eqs. (\ref{eq:force_constant_symmetry}) and (\ref{eq:dyn_mat}).

We conclude this subsection by emphasizing that the space group of the actual BaMnSb$_2$ crystal is symmorphic \cite{liu2021spin}; i.e. it does not contain the $G_{2y}$ operation. Thus, we will henceforth disregard $G_{2y}$, though it is possible to show that its presence would rule out the phonon helicity and the phonon Berry curvature.

\subsection{Effective dynamical matrices}

In this work, we are interested in long-wavelength (small-$q$) phonons. At $q= 0$, the dynamical matrix has two  degenerate zero-frequency acoustic modes with eigenvectors 
\begin{align}
\label{eq:ac0}
|A_{0, x}\rangle &=(1/\sqrt{2}) (1,0,1,0)^T \notag\\
|A_{0, y}\rangle &=(1/\sqrt{2}) (0,1,0,1)^T 
\end{align}
and two nondegenerate optical modes with eigenvectors
\begin{align}
\label{eq:op0}
|O_{0, x}\rangle &=(1/\sqrt{2}) (1,0,-1,0)^T \text{   ($A_1$ mode)} \notag\\
|O_{0, y}\rangle &=(1/\sqrt{2}) (0,1,0,-1)^T \text{   ($B_1$ mode)}.
\end{align}
At finite wave vectors, one must diagonalize $D({\bf q})$ to obtain the phonon polarizations. 
Yet, an approximate analytical approach can be pursued for small $q$ values (near the $\Gamma$ point of Fig. \ref{fig:phonondispersion}), where the optical and acoustic phonon branches are well separated from one another in frequency.
The approximation consists of decomposing the dynamical matrix in two decoupled $2 \times 2$ matrices,
$D({\bf q}) \simeq {\cal D}^a ({\bf q}) \oplus {\cal D}^o ({\bf q})$,
where
\begin{align}
\label{eq:eff_D}
{\cal D}^a ({\bf q}) &\equiv
\begin{pmatrix}
\langle A_{0, x} | D({\bf q}) |A_{0, x}\rangle & \langle A_{0, x} | D({\bf q}) |A_{0, y}\rangle\\
\langle A_{0, y} | D({\bf q}) |A_{0, x}\rangle & \langle A_{0, y} | D({\bf q}) |A_{0, y}\rangle
\end{pmatrix}\notag\\
{\cal D}^o ({\bf q})&\equiv
\begin{pmatrix}
\langle O_{0, x} | D({\bf q}) |O_{0, x}\rangle & \langle O_{0, x} | D({\bf q}) |O_{0, y}\rangle\\
\langle O_{0, y} | D({\bf q}) |O_{0, x}\rangle & \langle O_{0, y} | D ({\bf q})|O_{0, y}\rangle
\end{pmatrix}
\end{align}
are the effective acoustic and optical dynamical matrices. 
They can be written as
\begin{equation}
{\cal D}^{a/o} = d^{a/o}_0 \sigma_0+ \sum_{i=x,y,z} d^{a/o}_i \sigma_i,
\label{Eq:dynamcial_2by2}
\end{equation}
where $\sigma_i$ is the effective pseudospin in the $\{x, y\}$ basis.
Starting from the expressions for $D_{\alpha\beta}(\kappa,\kappa';{\bf q})$ in App. \ref{ap:dynamical} and doing a small $q$ approximation, we obtain
\begin{align}
\label{eq:ds}
d^{a(o)}_0 &\simeq \frac{1}{M}\left[ \gamma_+ \left( 1 \mp (1-\frac{q^2 a^2}{8})\right) + \gamma_0 \frac{q^2 a^2}{2} \right]\notag \\
d^{a(o)}_z &\simeq  \frac{1}{M}\left[ \gamma_- \delta a \left( 1 \mp (1-\frac{q^2  a^2}{8})  \right) + \gamma_0  a^2 \frac{q_x^2 - q_y^2}{2} \right]\notag\\
d^{a(o)}_x &\simeq \pm \frac{1}{M}\gamma_+ a^2 \frac{q_x q_y}{4}\notag \\
d^{a(o)}_y &= 0,
\end{align}
where $M$ is the mass of an Sb atom, $\gamma_\pm \equiv \gamma_2 \pm \gamma_1$ and we have assumed that the zig-zag distortion ($\delta a$) is small. The upper (lower) signs in Eq.~(\ref{eq:ds}) correspond to acoustic (optical) modes.

Symmetry wise, projecting Eq.~(\ref{eq:vosko}) onto the optical ($|O_{0, x}\rangle,  |O_{0, y}\rangle$) or acoustic ($|A_{0, x}\rangle,  |A_{0, y}\rangle$) subspaces, we get the constraint
\begin{equation}
\label{eq:C2sym}
\sigma_z {\cal D}^{o/a} (q_x,q_y) \sigma_z = {\cal D}^{o/a} (q_x,-q_y) 
\end{equation}
from the $C_{2x}$ or $\sigma_v(xz)$ symmetries, which is satisfied by Eq. (\ref{eq:ds}).

\subsection{Berry curvature}
\label{sec:BC_phonon1}

In this section, we calculate the Berry curvature for long-wavelength acoustic and optical phonons in BaMnSb$_2$, in the absence of the coupling to Dirac fermions. 
The first step is to obtain the effective phonon Hamiltonian. As shown in Ref.~\cite{liu2017model}, an effective Hamiltonian for phonons in the presence of time-reversal symmetry can be written as
\begin{equation}
\label{eq:heff}
{\cal H}_{\rm ph}({\bf q}) = {\cal D}^{1/2}({\bf q}). 
\end{equation}
There exist alternative expressions for the effective phonon Hamiltonian in the literature \cite{zhang2010topological, qin2012berry, sun2012current}, depending on whether the Hamiltonian is written in terms of the canonical or mechanical momenta.
All these expressions are unitarily equivalent when time-reversal symmetry is preserved.

Thus, the effective $2\times 2$  Hamiltonians for acoustic and optical phonons are simply the square roots of ${\cal D}^{a}$ and ${\cal D}^o$, respectively.
Omitting for brevity the superscripts $a$ and $o$, as well as the wave vector label ${\bf q}$, it is easy to show that
\begin{equation}
\label{eq:hph}
{\cal H}_{\rm ph} = {\cal D}^{1/2} =  V \mathbb{D}^{1/2} V^\dagger,
\end{equation}
where $V$ is a unitary matrix that diagonalizes ${\cal D}$,
\begin{equation}
\mathbb{D} = V^\dagger {\cal D} V 
=\begin{pmatrix}
d_0+d & 0 \\
0 & d_0-d
\end{pmatrix}
\equiv 
\begin{pmatrix}
\omega_1^2 & 0 \\
0 & \omega_2^2
\end{pmatrix}
\end{equation}
is the diagonalized effective dynamical matrix for the acoustic or optical modes, and
$d$ is the modulus of ${\bf d} = (d_x, d_y, d_z)$ listed in Eq.~(\ref{eq:ds}).
Using
$\mathbb{D}^{1/2} = {\rm diag}(\omega_1, \omega_2)$
and
\footnote{We make the gauge choice to place the phase factor $\exp(\pm i\phi)$ always in front of $\sin(\theta/2)$. We could equally well have chosen to place the phase factors always in front of $\cos(\theta/2)$. Either choice will ensure that $|1\rangle\langle 2|$  has no discontinuities as the wave vector is varied ($|1\rangle\langle 1|$ and $|2\rangle\langle 2|$ already have that property irrespective of the gauge choice).}
\begin{equation}
V = \begin{pmatrix}
\cos \frac{\theta}{2} & \sin \frac{\theta}{2} e^{-i \phi}\\
\sin \frac{\theta}{2} e^{i \phi} & - \cos \frac{\theta}{2} 
\end{pmatrix}
\label{Eq:V}
\end{equation}
with $\cos \theta = d_z/d$ and  $\tan \phi = d_y/d_x$, Eq. (\ref{eq:hph}) becomes
\begin{equation}
\label{eq:phoha}
{\cal H}_{\rm ph} = h_0 \sigma_0 +  \sum_{i=x,y,z} h_i \sigma_i,
\end{equation}
where
\begin{align}
\label{eq:h}
h_0 &= \frac{\omega_1 + \omega_2}{2} \notag \\
h_i &= \frac{\omega_1 - \omega_2}{2 d} d_i.
\end{align}
Here, $\sigma_i$ is once again the pseudospin in the $\{x,y\}$ basis.
A few remarks are in order:
(i) For the effective Hamiltonian to be defined, we should have $d_0 >d$, i.e. the eigenvalues of the dynamical matrix must be non-negative; (ii) the eigenvalues of the effective Hamiltonian are the phonon frequencies $\omega_1$ and $\omega_2$; (iii) 
the eigenvectors of ${\cal H}_{\rm ph}$ coincide with those of ${\cal D}$.

From Eq.~(\ref{eq:phoha}), the Berry curvature can be obtained using~\cite{sticlet2012geometrical}
\begin{equation}
\label{eq:BC}
\boldsymbol{\Omega}_\pm ({\bf q}) = \pm \hat{\bf z}\frac{1}{2 h^3} {\bf h}\cdot \left(\partial_{q_x}{\bf h} \times \partial_{q_y}{\bf h}\right),
\end{equation}
where ${\bf h}=(h_x,h_y,h_z)$, $\pm$ corresponds to the two acoustic or optical in-plane phonon modes, and $h$ is the modulus of ${\bf h}$
 \footnote{In the presence of time-reversal symmetry, the eigenvectors of ${\cal H}_{\rm ph}$ coincide with those of ${\cal D}$. Consequently, the Berry curvature can be calculated directly from the dynamical matrix without having to deal with the Hamiltonian. However, in the end of the paper we will consider the case of broken time-reversal symmetry. In that case, the effective phonon Hamiltonian is no longer given by the square root of the dynamical matrix, and its eigenvectors no longer coincide with those of the dynamical matrix. In that more general case, we can still obtain the phonon Berry curvature from Eq.~(\ref{eq:BC}).}.

Combining Eqs. (\ref{eq:h}) and (\ref{eq:BC}) immediately gives $\boldsymbol{\Omega}_\pm ({\bf q}) =0$.
The mathematical reason for it is that $h_y=0$ in Eq. (\ref{eq:h}), because $d_y=0$ in Eq. (\ref{eq:ds}).
Indeed,  $h_x$, $h_y$ and $h_z$ must all be nonzero for a Berry curvature to emerge.
Thus, {\it bare} phonons in BaMnSb$_2$ do not have any Berry curvature. 

From more physical viewpoint, the underlying reason for the vanishing of the Berry curvature is that our force constant model obeys Eq. (\ref{eq:force_constant_symmetry}). This condition leads to $h_y=0$. To see that, it suffices to show that the off-diagonal element of ${\cal D}^{o,a}$, denoted here as ${\cal D}^{o,a}_{xy}$, is purely real.
From Eqs. (\ref{eq:dyn}) and (\ref{eq:eff_D}), we have
\begin{align}
\label{eq:dofdiag}
{\cal D}^{o,a}_{xy}({\bf q}) =\frac{1}{2}&\left[\mp D_{xy}(1,2) \mp D^*_{yx}(1,2)\right.\nonumber\\
&\left. +D_{xy}(1,1)+D_{xy}(2,2)\right].
\end{align}
If Eq. (\ref{eq:force_constant_symmetry}) is obeyed, then it follows from Eq. (\ref{eq:dyn_mat}) that $D_{xy}(1,2)=D_{yx}(1,2)$. Consequently, the first line of Eq. (\ref{eq:dofdiag}) is purely real.
The second line is likewise purely real, because Eq.  (\ref{eq:force_constant_symmetry}) implies $D_{xy}(\kappa,\kappa)=D_{yx}(\kappa,\kappa)$, whereas hermiticity of the dynamical matrix in Eq. (\ref{eq:dyn}) imposes $D_{yx}(\kappa,\kappa)=D_{xy}(\kappa,\kappa)^*$. Thus, $D_{xy}(\kappa,\kappa)$ must be a real number. 

In sum, force constants satisfying Eq. (\ref{eq:force_constant_symmetry})  preclude a phonon Berry curvature.
As we show next, the coupling between phonons and Dirac fermions changes this state of affairs, by effectively providing force constants $\Phi_{\alpha\beta}(l \kappa,l^\prime \kappa^\prime)$ that are not symmetric under exchange of $\alpha$ and $\beta$.

\section{Electron-phonon interaction}
\label{sec:e-ph}

Because the low-energy electronic states originate mainly from Sb layers, they couple significantly to the vibrations of Sb atoms. The unscreened electron-phonon interaction Hamiltonian reads \cite{rinkel2017signatures,saha2015detecting}
\begin{equation}
{\cal H}_{\rm e-ph} =  \sum_{{\bf k},{\bf q}} \sum_{\lambda,s,\alpha,\beta}  g^{\lambda,s}_{\alpha,\beta} ({\bf q})  u_{{\bf q},\lambda} c^\dagger_{s,\alpha,{\bf k}} c_{s,\beta,{\bf k} - {\bf q}},
\label{Eq:Electron-phonon}
\end{equation} 
where $u_{{\bf q},\lambda} = (b_{{\bf q},\lambda} + b^\dagger_{-{\bf q},\lambda})/\sqrt{2 M_{\rm uc} \omega_{{\bf q},\lambda}}$ is the phonon displacement operator, $M_{\rm uc}$ is the atomic mass of the unit cell (in our simplified model, we have $M_{\rm uc}=2 M$),
$b_{{\bf q},\lambda}$ is an operator that annihilates a phonon mode $\lambda$ with momentum ${\bf q}$, $\alpha,\beta$ represent electronic $p$-orbitals and  $g^{\lambda,s}_{\alpha,\beta}$ is the electron-phonon matrix element. 
Only electronic valley-preserving (i.e. long-wavelength) phonons are considered, such that we may write ${\cal H}_{\rm e-ph}=\sum_s{\cal H}^s_{\rm e-ph}$.
The $2\times 2$ electron-phonon coupling matrix at valley $s$ satisfies the hermiticity condition 
\begin{equation}
\label{eq:herm}
g^{\lambda,s} ({\bf q})^\dagger = g^{\lambda,s} (-{\bf q})
\end{equation}
and the time-reversal symmetry condition
\begin{equation}
\label{eq:trev}
 g^{\lambda,s} ({\bf q}) = [g^{\lambda,-s} (-{\bf q}) ]^*.
 \end{equation}
The dependence of $g^{\lambda,s}_{\alpha,\beta}$ on ${\bf k}$ (the electronic wave vector measured from a given valley $s$) is neglected, but its valley-dependence is considered through the label $s$.
The form of the allowed terms in $g^{\lambda,s}_{\alpha,\beta}$ can be determined from symmetry analysis \cite{hu2021phonon}.

In this study, we focus only on the interaction between the low-energy Dirac fermions and the phonons. 
Moreover, we concentrate on the deformation potential coupling for optical and acoustic phonons, thereby neglecting the Fr\"olich and the piezoelectric interactions \cite{mahan2013many, vogl1976microscopic}.

\subsection{Optical deformation potential}

The optical deformation potential matrix $g^{\lambda,s}$ can be approximated to be independent of ${\bf q}$ for long-wavelength optical phonons. 
This approximation is enabled by the fact that the two $q=0$ optical modes of interest, $|O_{0, x}\rangle$ ($A_1$) and $|O_{0, y}\rangle$ ($B_1$) in Eq.~(\ref{eq:op0}), are not degenerate. 
The invariance of ${\cal H}_{\rm e-ph}^s$ under the combined symmetry $C_{2x} {\cal T}$ then implies \cite{hu2021phonon}
\begin{equation}
\label{eq:tot}
\tau_z (g^{\lambda,s})^* \tau_z = x_\lambda g^{\lambda,s},
\end{equation}
where Eq.~(\ref{eq:op0}) has been used through
\begin{equation}
(C_{2x} {\cal T}) u_{{\bf 0}, \lambda} (C_{2x} {\cal T})^{-1} = x_\lambda u_{{\bf 0}, \lambda}
\end{equation}
with $x_{A_1} = +1 $ and $x_{B_1} = -1 $.
For a given valley $s\in\{+,-\}$, Eqs. (\ref{eq:tot}), (\ref{eq:trev}) and (\ref{eq:herm}) lead to
\begin{align}
\label{eq:g0s}
g^{A_1,s} &= g_0^o \tau_0 + s g_y^o \tau_y + g_z^o \tau_z \nonumber \\
g^{B_1,s} &= g_x^o \tau_x,
\end{align}
where $g^o_{0,1,2,3}$ are real ${\bf q}$-independent numbers.

\subsection{Acoustic deformation potential}
Acoustic phonons become degenerate at $q\to 0$. For that reason,  as soon as $q$ becomes infinitesimally nonzero, $|A_{0,x}\rangle$ and $A_{0, y}\rangle$ are no longer the eigenmodes of the dynamical matrix for a generic $\hat{\bf q}$  (not even approximately so); the eigenmodes must instead be obtained from diagonalizing ${\cal D}^a$.
For simplicity, we will approximate the two in-plane acoustic modes in the $q\to 0$ limit as longitudinal (LA) and transverse (TA), with polarization vectors
\begin{align}
\label{eq:pLT}
{\bf p}_{{\bf {\hat q}}, LA} &= \hat{\bf q}\notag\\
{\bf p}_{{\bf {\hat q}}, TA} &= \hat{\bf q} \times \hat{\bf z}.
\end{align}
The exact polarizations, obtained by diagonalizing ${\cal D}^a$ in the $q\to 0$ regime, tend to Eq.~(\ref{eq:pLT}) under the assumption of isotropy (neglect of $\delta a$ and assumption of $\gamma_0 = \gamma_+/4$). For qualitative purposes, this approximation is satisfactory when the zig-zag distortion is small. For analytical convenience, we will continue to use Eq.~(\ref{eq:pLT}).

From the acoustic sum rule, we know that $g^{\lambda, s} ({\bf q})$ must vanish when $q\to 0$.
Taylor-expanding the microscopic expression for the electron-phonon matrix elements (found e.g. in the Supplemental Material of \cite{hu2021phonon}) in powers of $q$, the leading order deformation potential term is of the form
\begin{equation}
\label{eq:adef0}
g^{\lambda, s} ({\bf q}) =  \sum_{\nu, \nu'}\Xi^{\lambda, s}_{\nu, \nu'} q_\nu {\bf p}_{\hat {\bf q}\lambda, \nu'},
\end{equation}
where $\nu,\nu'\in\{x,y\}$ and $\Xi^{\lambda, s}_{\nu, \nu'}$ is a ${\bf q}$-independent $2\times 2$ matrix.

If we assume isotropy, the only two vectors governing $g^{\lambda, s} ({\bf q})$  are ${\bf q}$ and ${\bf p}_{{\bf q}, \lambda}$.
Since the electron-phonon matrix elements are scalars, they must involve either ${\bf q}\cdot{\bf p}_{\hat {\bf q} \lambda}$ or $({\bf q} \times\hat{\bf z})\cdot{\bf p}_{\hat {\bf q} \lambda}$ (the only two scalars that can be formed from the aforementioned two vectors, $\hat{\bf z}$ being the unit vector normal to the Sb planes).
Then, Eq.~(\ref{eq:adef0}) can be simplified to 
\begin{equation}
g^{\lambda, s} ({\bf q}) =  \Xi^{LA, s} {\bf q}\cdot{\bf p}_{\hat {\bf q} \lambda} + \Xi^{TA, s} ({\bf q} \times\hat{\bf z})\cdot{\bf p}_{\hat {\bf q} \lambda}, 
\end{equation}
where the first (second) term in the right hand side only appears for the longitudinal (transverse) mode.
As a result, 
\begin{equation}
\label{eq:adef}
g^{\lambda, s} ({\bf q}) =  \Xi^{\lambda, s} q,
\end{equation}
where $\Xi^{\lambda, s}$ is a $q$-independent $2\times 2$ matrix.
For similar approximations in related systems, see e.g. Refs. \cite{kaasbjerg2012phonon, song2013transport}.

In order to identify the nonzero elements of the deformation potential matrix $\Xi^{\lambda, s}$, we use symmetry arguments.
First, we recognize that 
\begin{equation}
(C_{2x} {\cal T}) {\bf p}_{{\bf {\hat q}}, \lambda} (C_{2x} {\cal T})^{-1} = x_\lambda {\bf p}_{(C_{2x}{\cal T}) {\bf {\hat q}}, \lambda},
\end{equation}
with $x_{LA} = -1$ and $x_{TA} = +1$.
Using the invariance of ${\cal H}^s_{\rm e-ph}$ under $C_{2x} {\cal T}$ and keeping in mind that $u_{{\bf q},\lambda}$ transforms in the same way as ${\bf p}_{{\bf q},\lambda}$ under symmetry operations, we have

\begin{equation}
\tau_z (g^{\lambda,s}(q_x,q_y))^* \tau_z =  x_\lambda g^{\lambda,s}(-q_x,q_y).
\end{equation}
Combining this with Eqs. (\ref{eq:herm}) and (\ref{eq:trev}), we get
\begin{align}
\label{eq:adef2}
\Xi^{LA,s} &= g_x^a \tau_x  \nonumber \\
\Xi^{TA,s} &= g_0^a \tau_0 + s g_y^a\tau_y + g_z^a \tau_z,
\end{align}
where $g^a_{0,1,2,3}$ are real ${\bf q}$-independent numbers.
Note that the couplings $g^o$ and $g^a$ do not have the same units (the former have dimensions of energy/length, while the latter have dimensions of energy).

\section{Dressed in-plane phonons}
\label{sec:dressed_phonons}

\subsection{Dynamical matrix}

The electron-phonon interaction of Sec. \ref{sec:e-ph} modifies the phonon frequencies and eigenvectors calculated in Sec. \ref{sec:bare_phonons}.
In this section, we will  find out how the effective $2\times 2$ dynamical matrices ${\cal D}^{a}$ and ${\cal D}^o$ for the acoustic and optical modes are modified in the presence of electron-phonon interactions.
We will neglect the hybridization between bare optical and acoustic modes caused by electrons, instead concentrating on the changes occurring within the optical and acoustic mode subspaces.
One convenient way to identify these changes is from the effective action for optical or acoustic phonons, obtained by integrating out the Dirac fermions. 
We have detailed this procedure in Ref.~\cite{parlak2023detection}.
Here, we simply quote the outcome with relevant details relegated in App. \ref{ap:action},
\begin{align}
\label{Eq:seffAB_o}
S_{\rm eff}&=\frac{M_{\rm uc}}{2} \sum_{{\bf q}, q_m} \left(u^*_{{\bf q}, 1} (q_m), u^*_{{\bf q}, 2}(q_m)\right)\times\notag\\
&\left(\begin{array}{cc}  q_m^2 + \omega_1^2 + \Sigma_{11} & \Sigma_{12} \\ \Sigma_{21}&  q_m^2 + \omega_{2}^2+\Sigma_{22}\end{array}\right) \left(\begin{array}{c}u_{{\bf q}, 1}(q_m) \\ u_{{\bf q}, 2}(q_m) \end{array}\right),
\end{align}
where $q_m$ a bosonic Matsubara frequency and $\Sigma_{\lambda \lambda'}$ is an element of the phonon self-energy (dependent on ${\bf q}$ and $q_m$) caused by its coupling to electrons. The subscripts $1$ and $2$ label the two in-plane, bare acoustic or optical phonons: in the case of optical phonons, $1$ ($2$) corresponds to the $A_1$ ($B_1$) mode; in the case of acoustic phonons, $1$ ($2$) corresponds to the longitudinal $L$ (transverse $T$) mode.

We will be interested in the low-temperature regime, and in phonon frequencies lower than the electronic bandgap. 
In that regime, $\Sigma_{\lambda\lambda}$  is purely real and can be simply absorbed into $\omega_{\lambda}$  (cf. App. \ref{ap:action}).
Concerning the off-diagonal self-energies, their low-frequency and long-wavelength expansion reads
\begin{equation}
\label{eq:AB}
\Sigma_{AB}= -\Sigma_{BA} \simeq  -\frac{g_0^o g_x^o {\cal S}}{M_{\rm uc}} \frac{i q_y}{2\pi v_0} {\rm sgn} (m_0)
\end{equation}
for the case of optical phonons.
 For acoustic phonons, the corresponding expressions are 
\begin{equation}
\label{eq:LT}
\Sigma_{TL} = -\Sigma_{LT} \simeq  -\frac{g_0^a g_x^a {\cal S}  }{M_{\rm uc}} \frac{i q^2 q_y}{2\pi v_0} {\rm sgn} (m_0).
\end{equation}
It must be mentioned that the phonon self-energies are independent of the system area ${\cal S}$, since the couplings $g$ scale as $1/\sqrt{\cal S}$.
For example, in the deformation potential approximation \cite{kaasbjerg2012phonon} that we will use in the numerical estimates, 
\begin{equation}
g^2\sim d^2 \frac{{\cal S}_{\rm uc}}{{\cal S}},
\end{equation}
 where $d$ is the deformation potential (in units of energy/length for the optical couplings $g^o$ and in units of energy for the acoustic couplings $g^a$) and ${\cal S}_{\rm uc}$ is the unit cell area.

In Eqs.~(\ref{eq:AB}) and (\ref{eq:LT}), the valley electronic Chern number ${\rm sgn}(m_0)$ enters directly.
The off-diagonal self-energies are static to leading order (i.e. they are finite at $q_m\to 0$).
In addition, the fact that these terms are imaginary renders the dressed phonons helical \cite{hu2021phonon}, with an angular momentum along the $z$ direction that inverts sign under $q_y\to -q_y$.  
Finally, the bare values of  $g_0^o g_x^o$ and $g_0^a g_x^a$ must be divided by the dielectric constant $\epsilon$ in order to account for electron-electron interactions in the random phase approximation. 

We reiterate that Eqs.~(\ref{eq:AB}) and (\ref{eq:LT}) are valid when the phonon frequency and the system's temperature are small compared to the energy gap $2 |m_0|$ of the insulator (a situation that can be relevant to BaMnSb$_2$, as indicated in  App. \ref{ap:action}). If $|m_0|$ becomes comparable to or smaller than the phonon frequency or the temperature, there will be additional terms in the phonon self-energy, aside from the topological terms (\ref{eq:AB}) and (\ref{eq:LT}).  These extra terms will depend on the details of the electronic band structure (e.g. the magnitudes of the Dirac mass $m_0$ and the Dirac velocity $v_0$), as well as on the phonon frequency. In real frequencies, the diagonal self-energies $\Sigma_{11}$ and $\Sigma_{22}$  will acquire imaginary parts, associated to the phonon decay into particle-hole pairs, and the off-diagonal phonon self-energies $\Sigma_{12}$ and $\Sigma_{21}$ will acquire real parts. A detailed study of these terms is beyond the scope of our work, but should not change the main message from our study, namely that there is an electronic topological contribution to the Berry curvature of long-wavelength phonons.

The equation of motion for the dressed phonons can be obtained by minimizing the effective action:
\begin{equation}
\frac{\delta S_{\rm eff}}{\delta u_{{\bf q},\lambda}(q_m)} = 0
\end{equation}
for $\lambda=1,2$. This yields
\begin{equation}
\label{eq:eom}
(i q_m)^2\left(\begin{array}{c}u_{{\bf q}, 1}(q_m) \\ u_{{\bf q}, 2}(q_m) \end{array}\right)= \tilde{\mathbb{D}}({\bf q}) \left(\begin{array}{c}u_{{\bf q}, 1}(q_m) \\ u_{{\bf q}, 2}(q_m) \end{array}\right),
\end{equation}
where 
\begin{equation}
\tilde{\mathbb{D}}({\bf q}) = \left(\begin{array}{cc} \omega_1^2 + \Sigma_{11} & \Sigma_{12} \\ -\Sigma_{12} & \omega_2^2+\Sigma_{22}\end{array}\right)
\end{equation} 
is the renormalized effective dynamical matrix for in-plane optical ($1=A, 2=B$) or in-plane acoustic ($1=L, 2=T$) modes.

It is important to note that this dynamical matrix is written in the eigenbasis $\{1, 2\}$ of the bare phonons,
whereas the one obtained  in Eq. (\ref{Eq:dynamcial_2by2}) is in the $\{x, y \}$ basis. 
We therefore carry out a unitary transformation
$\tilde{\cal D} = W \tilde{\mathbb{D}} W^\dagger$, where $W$ is given by Eq.~(\ref{Eq:V}) for the case of optical phonons and 
\begin{equation}
W=\begin{pmatrix} \hat{\bf q}\cdot\hat{\bf x} & -\hat{\bf q}\cdot\hat{\bf y}\\
\hat{\bf q}\cdot\hat{\bf y} & \hat{\bf q}\cdot\hat{\bf x} 
\end{pmatrix}
\end{equation} 
for the case of acoustic phonons. After performing the transformation, we have the  effective dynamical matrices for the dressed acoustic and optical modes in the $\{x, y \}$ basis:
\begin{equation}
\label{eq:ren_dyn}
\tilde{\cal D}^{a/o} = \tilde{d}^{a/o}_0 \sigma_0+ \sum_{i=x,y,z} \tilde{d}^{a/o}_i \sigma_i,
\end{equation}
where 
\begin{align}
\tilde{d}^{o(a)}_0 &= d_0^{o(a)}+\frac{\Sigma_{11}+\Sigma_{22}}{2}\notag\\
\tilde{d}^{o(a)}_z &= d_z^{o(a)}\left(1+\frac{\Sigma_{11}-\Sigma_{22}}{2d^{o(a)}}\right)\notag\\
\tilde{d}^{o(a)}_x &= d_x^{o(a)}\left(1+ \frac{\Sigma_{11}-\Sigma_{22}}{2 d^{o(a)}}\right) \notag\\
\tilde{d}^{o(a)}_y &=  \pm {\rm Im} \Sigma_{12}.
\label{Eq:dynamical_2by2_final}
\end{align}
The diagonal phonon self-energy terms $\Sigma_{11}$ and $\Sigma_{22}$ simply renormalize the spring constants appearing in the dynamical matrix of the bare phonons. 
In contrast, the imaginary off-diagonal self-energy $\Sigma_{12}$ leads to new physical effects that cannot be accounted for using the bare  spring constants of Sec. \ref{sec:bare_phonons}. 

In density functional theory \cite{giustino2017electron, coh2023classification}, the imaginary off-diagonal phonon self-energy and the corresponding phonon helicity should be obtainable within the adiabatic (Born-Oppenheimer) approximation with dressed force constants, because said self-energy has been computed in the static ($q_m\to 0$) limit.
Yet, the various empirical force-constant models we have used to obtain the dynamical matrix of Sec. \ref{sec:bare_phonons} fail to predict phonon helicity.  
Indeed, as we proved in Sec. \ref{sec:bare_phonons}, $\tilde{d}_y=0$ if the force constant matrix obeys Eq. (\ref{eq:force_constant_symmetry}). It is therefore clear that the coupling of Dirac fermions to lattice vibrations violates Eq. (\ref{eq:force_constant_symmetry})  by contributing terms to the renormalized force constants that are {\em asymmetric} under the exchange of $\alpha$ and $\beta$. Such terms, proportional to the electronic valley Chern number, generate phonon helicity.
 
Another key aspect of the off-diagonal phonon self-energy is that it generates a phonon Berry curvature. In Sec. \ref{sec:BC_phonon1}, we observed that the Berry curvature of bare phonons was zero due to $d_y=0$ in Eq.~(\ref{Eq:dynamcial_2by2}). Now that  $\tilde{d}_y\neq 0$, we expect the dressed phonons to have a nonzero Berry curvature.

\subsection{Berry curvature}
\label{Sec:BC_phonon2} 
\begin{figure}
\includegraphics[width=0.48\textwidth]{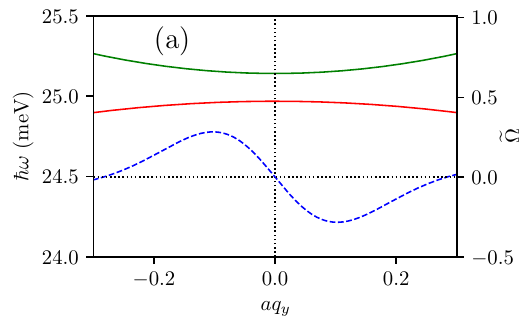}
\includegraphics[width=0.48\textwidth]{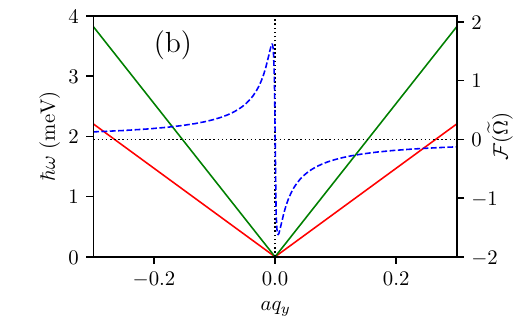}
\caption{Phonon dispersion and Berry curvature for optical (a) and acoustic (b) modes around the $\Gamma$ point. The phonon branches calculated using Eq.~(\ref{Eq:dynamcial_2by2}) are shown in solid red and green lines. The electron-phonon interaction parameters are defined as $g_0^o g_x^o {\cal S} = {d^o}^2 {\cal S}_{\rm uc}$ and $g_0^a g_x^a  {\cal S} = {d^a}^2 {\cal S}_{\rm uc}$ with a deformation potential $d^o=1\, {\rm eV/\AA}$ and $d^a =  5 \, {\rm eV}$ for optical and acoustic modes, respectively, and a unit cell area ${\cal S}_{uc}=a^2$. The dashed blue line shows the dimensionless Berry curvature $\widetilde{\Omega} = \Omega_-/ {\cal S}_{\rm uc}$ for the upper phonon branches (green lines). Lower phonon branches (red lines) have the opposite Berry curvature. Note the different scales for left and right axes for the two figures. For acoustic phonons, ${\cal F}(\widetilde{\Omega}) = {\rm sgn}(\widetilde{\Omega}) \ln(1+|\widetilde{\Omega}|)$
is plotted for better visibility. Also, $a q_x = 0.001$ is taken in $\Omega$ calculation for acoustic modes to circumvent the $q = 0$ divergence. Here, we have taken the Dirac velocity $v_0 = 10^6 \, {\rm m/s}$, the lattice constant $a=5 \AA$~\cite{sakai2020bulk}, and other parameters are same as in Fig.~\ref{fig:phonondispersion}.
}
\label{fig:BC}
\end{figure}
Let us again calculate the Berry curvature for phonons, as in Sec.~\ref{sec:BC_phonon1}, but this time using the effective Hamiltonian of dressed phonons. For the long-wavelength optical modes, we obtain
\begin{equation}
\label{eq:Berry_optical_dressed}
\boldsymbol{\Omega}_\pm({\bf q}) \simeq \pm \hat{\bf z} \frac{M \gamma_+ a^2 {\rm sign}(\gamma_- \delta a)}{32 \gamma_-^2 \delta a^2}{\rm Im}\Sigma_{AB},
\end{equation}
where we have expanded in powers of $q$ first and afterwards assumed that $ \gamma_-\delta a\ll \gamma_+$ (weak zig-zag distortion). 
  For the long-wavelength acoustic modes, we get
 \begin{equation}
 \label{eq:Berry_acoustic_dressed}
\boldsymbol{\Omega}_\pm({\bf q}) \simeq \pm \hat{\bf z} \frac{2 M }{\gamma_0 q^4 a^2}{\rm Im}\Sigma_{LT},
\end{equation}
where we have adopted the isotropic approximation from the onset (neglecting $\delta a$ and assuming $\gamma_+ = 4 \gamma_0$).

These results are illustrated in Fig.~\ref{fig:BC}. The main conclusion from these expressions is that Berry curvature of the dressed optical and acoustic phonons is proportional to the phonon helicity and thus to the electronic valley Chern number.
In some sense, phonons have {\it inherited} the Berry curvature from electrons through the electron-phonon interactions. This is the main result from the present work.

In Ref. \cite{hu2021phonon}, the electronic valley Chern number was shown to generate a phonon angular momentum. Therefore, it is worthwhile to look at its connection with the phonon Berry curvature. From Ref. \cite{hu2021phonon}, one can derive the following expression for the frequency- and momentum-resolved phonon angular momentum of long-wavelength phonons:
\begin{equation}
{\bf L} \simeq \hat{\bf z}  \frac{2 {\rm Im}\Sigma_{12}}{\omega_1^2-\omega_2^2}\left[\delta(\omega^2-\omega_1^2)-(1\leftrightarrow 2)\right],
\end{equation}
where $\omega$ is the real frequency obtained by analytic continuation $q_m \rightarrow - i\omega + 0^+$. The Dirac delta functions are broadened into Lorentzians in the presence of phonon decay processes that are not modelled in the present work. Thus, both the Berry curvature and angular momentum of the dressed phonons are  proportional to the electronic valley Chern number. Also, they have opposite directions for the phonon branches $1$ and $2$.

We notice in passing that, in the limit $q\to 0$, the Berry curvature of optical phonons vanishes. For acoustic phonons, the Berry curvature has a divergence and a discontinuity at $q=0$, because it scales as $q_y/q^2$. This singularity can be associated to the inevitable degeneracy of the acoustic phonon bands at $q\to 0$. Analogously, the electronic Berry curvature is known to diverge at electron band-touching points. However, in the case of acoustic phonons, this singularity should not have any observable consequences because $q=0$ acoustic phonons describe static, rigid translations of the crystal. In addition, physical quantities such as thermal conductivity involve integration over phonon wave vectors and sums over phonon branches; the integration phase space proportional to $q$ will then compensate the $1/q$ behavior of the acoustic phonon Berry curvature, whereas the sum over the two nearly degenerate acoustic branches will tend to have a cancelling effect on their Berry curvatures.

More generally, the phonon Berry curvature obeys the relations ${\boldsymbol \Omega}_\pm({\bf q}) = -\boldsymbol{\Omega}_\pm(-{\bf q})$, $\boldsymbol{\Omega}_\pm(q_x, q_y) = -\boldsymbol{\Omega}_\pm(q_x, -q_y)$ and $\boldsymbol{\Omega}_\pm(q_x, q_y)=\boldsymbol{\Omega}_\pm(-q_x, q_y)$, which are consistent with the ${\cal T}$, $C_{2x}$ and $C_{2x}{\cal T}$ symmetries of the crystal, respectively. 
Consequently, the average of the Berry curvature over ${\bf q}$-space vanishes. 
The locally nonzero phonon Berry curvature could nevertheless have an impact on observable quantities 
\footnote{For example, the nonlinear thermal Hall effect might be sensitive to the phonon Berry curvature in ${\cal T}$-preserving systems; see e.g. Ref. \cite{varshney2023intrinsic}. We will not explore such possibility in the present work since the effect may be hardly measurable, as the linear phonon
thermal Hall conductivity that is being measured nowadays is already small.}.
In the next section, we will see that a ${\cal T}$-breaking perturbation leads to a nonzero ${\bf q}$-space average of the phonon Berry curvature, thereby producing a thermal Hall effect of phonons. 

\section{Effect of a time-reversal breaking perturbation}
\label{sec:TR_breaking}

In this section, we revisit the Berry curvature for optical and acoustic modes in the presence of a ${\cal T}$-breaking perturbation. 
For simplicity, we will limit ourselves to the effect of a spatially uniform ferromagnetic exchange field $B$ along the $z$ direction, whose effect on the electronic structure can be captured through a valley Zeeman splitting.
The same qualitative effect would be expected if we replaced the exchange field by a magnetic field and neglected the orbital effect (Landau-level splitting) due to the field.

In order to find out how $B$ influences the phonon Berry curvature, we need to determine how it enters the phonon effective Hamiltonian.
We will do so through a mix of phenomenological (symmetry) and microscopic arguments. 

According to \citep{liu2017model} (see the Supplemental Material in particular), the phonon effective Hamiltonian reads
\begin{equation}
\label{eq:HDG}
{\cal H}_{\rm ph}({\bf q}) = {\cal D}^{1/2}({\bf q}) - i {\cal G} ({\bf q}),
\end{equation}
where ${\cal G} ({\bf q})$ is the additional term arising out of ${\cal T}$-breaking.
The field $B$ does two things to Eq.~(\ref{eq:HDG}). First, it modifies the dynamical matrix; second, it leads to ${\cal G} ({\bf q})\neq 0$.
We will consider these two effects separately.

\subsection{Dynamical matrix}

$B\neq 0$ breaks the $C_{2x}$, $\sigma_v(xy)$, and ${\cal T}$ symmetries of BaMnSb$_2$. 
Accordingly, the counterparts of Eqs. (\ref{eq:Tsym}) and (\ref{eq:C2sym}) read
\begin{align}
{\cal D}^* ({\bf q},B) &= {\cal D}(-{\bf q},-B)\notag\\
\sigma_z {\cal D} (q_x,q_y,B) \sigma_z &= {\cal D} (q_x,-q_y,-B).
\label{Eq:Modified_constraints}
\end{align} 
The hermiticity condition ${\cal D}^\dagger ({\bf q}, B) = {\cal D}({\bf q}, B)$ holds unchanged.

For weak ${\cal T}$-breaking perturbations, we expand  the dynamical matrix to the leading order in $B$,
\begin{equation}
\label{eq:DB}
{\cal D}({\bf q},B) \simeq  {\cal D}^0({\bf q}) + \delta {\cal D}({\bf q}) B,
\end{equation}
where ${\cal D}^0({\bf q})$ is the dynamical matrix at $B=0$ as in Eq. (\ref{Eq:dynamical_2by2_final}).
Owing to the hermiticity condition, we can express 
\begin{equation}
\label{eq:little_a}
\delta {\cal D}({\bf q}) = \sum_{i} a_i ({\bf q}) \sigma_i
\end{equation}
in terms of Pauli matrices  ($i=0,x,y,z$), $a_i ({\bf q})$ being a real and $B-$independent function of ${\bf q}$. To obey Eq.~(\ref{Eq:Modified_constraints}), $a_0$ and $a_z$ must be odd functions of $q_y$ and even functions of $q_x$; $a_x$ must be an odd function of $q_x$ and an even function of $q_y$; $a_y$ must be an even function of both $q_x$ and $q_y$. 
 
Thus far we have made symmetry-based arguments. From a microscopic point of view, one can calculate the phonon self-energy to first order in the magnetic field, in the static ($q_m\to 0$) limit, so as to obtain the coefficients $a_i$. This calculation shows that, in the (rotationally symmetric) Dirac fermion model adopted in Eq. (\ref{eq:hek}), $\delta {\cal D}({\bf q}) =0$ (cf. App.~\ref{sec:delta_self}). We will henceforth neglect $\delta {\cal D}({\bf q})$, notwithstanding that there could be contributions to $\delta {\cal D}({\bf q})$ coming from high-energy (non-Dirac) electronic states. 
 
\subsection{Matrix ${\cal G} ({\bf q})$}

There are different physical mechanisms that can give raise to ${\cal G} ({\bf q})\neq 0$ in Eq.~(\ref{eq:HDG}) when $B\neq 0$ \cite{liu2017model}.
One mechanism responsible for ${\cal G}$ is the Raman spin-lattice interaction (see e.g. Ref. \cite{zhang2010topological} and references therein).
 This interaction contributes a term, proportional to the electronic magnetization,  to the phonon Hamiltonian. In our approximate low energy-model of BaMnSb$_2$, where the valleys are fully spin-polarized (with opposite spin polarization at the two valleys), a spatially uniform $B$ does not induce an electronic magnetization insofar as temperature is low compared to the energy gap and the chemical potential remains inside the gap \footnote{The spin susceptibility in this regime originates from interband matrix elements of the spin operator and is proportional to the square of the wave vector of the applied perturbation. It therefore vanishes for a spatially uniform $B$.}. We will thus neglect the Raman spin-lattice interaction. 

A second source of ${\cal G} ({\bf q})$ originates from the change in phonon self-energy due to the Zeeman splitting. When the ${\cal T}$-symmetry is broken by an exchange field, the electronic bands undergo a Zeeman shift. On one hand, the bands in one valley get shifted in energy with respect to the bands on the other valley, as the two valleys are oppositely spin-polarized. On the other hand, the difference in the $g-$factors between conduction and valence bands leads to a difference $2\zeta B$ in the energy gaps of the two valleys, where $\zeta$ is a coefficient proportional to the difference in the $g-$factors. This effect is known as the valley Zeeman splitting \cite{qi2015giant}. Due to it, the phonon self-energy is modified (cf. App.~\ref{sec:delta_self} for details).
As a result,  to first order in $B$ and in the long-wavelength approximation, the right hand side of the equation of motion in Eq.~(\ref{eq:eom}) acquires an extra term
\begin{equation}
\label{eq:delta_eom}
q_m\begin{pmatrix} \eta& \tilde{\eta}_s+ \tilde{\eta}_a \\\tilde{\eta}_s-\tilde{\eta}_a & 0\end{pmatrix}\begin{pmatrix} u_{{\bf q}, 1}(q_m) \\ u_{{\bf q}, 2} (q_m)\end{pmatrix},
\end{equation}
where
\begin{align}
\eta&=\frac{i q_y{\cal S} \zeta  B }{6 \pi M_{\rm uc}v_0 m_0^2} \times \left\{\begin{array}{cc} q^2 g^a_z g^a_0 & \text{(acoustic phonons)}\\
g^o_z g^o_0  & \text{(optical phonons)}\end{array}\right.\notag\\
\tilde{\eta}_s&=\frac{i q_x{\cal S} \zeta  B }{6\pi M_{\rm uc} v_0 m_0^2} \times \left\{\begin{array}{cc} q^2 g^a_x g^a_0 & \text{(acoustic ph.)}\\
g^o_x g^o_0  & \text{(optical ph.)}\end{array}\right.\notag\\
\tilde{\eta}_a&=\frac{q^2 {\cal S} \zeta B}{12\pi M_{\rm uc} |m_0|^3} \times \left\{\begin{array}{cc} q^2 g^a_x g^a_z & \text{(acoustic ph.)}\\
g^o_x g^o_z  & \text{(optical ph.)}\end{array}\right..
\end{align}
An important attribute of Eq. (\ref{eq:delta_eom}) is that is that it is linear in the phonon frequency; hence, it cannot be absorbed in a renormalized dynamical matrix.
Instead, it leads to the matrix ${\cal G} ({\bf q})$ of the effective phonon Hamiltonian. Comparing our equation of motion to that of Ref.~\cite{liu2017model}, we identify 
\begin{equation}
\mathbb{G}=\frac{1}{2} \begin{pmatrix} \eta & \tilde{\eta}_s+ \tilde{\eta}_a\\\tilde{\eta}_s- \tilde{\eta}_a & 0\end{pmatrix}
\end{equation}
in the $\{1,2\}$ basis of the bare phonons and thus
\begin{equation}
\label{eq:gscript}
{\cal G} ({\bf q}) = W \mathbb{G} W^\dagger 
\end{equation}
in the $\{x,y\}$ basis, where the unitary matrix $W$ has been discussed below Eq.~(\ref{eq:eom}).
Replacing Eq.~(\ref{eq:gscript}) in Eq.~(\ref{eq:HDG}) gives
\begin{align}
i{\cal G} &=i\frac{\eta}{2 d} \left(d \sigma_0 +d_x^o \sigma_x  + d_z^o \sigma_z \right)
\notag\\
&+i\frac{\tilde{\eta}_s}{d}\left(d_x^o\sigma_z - d_z^o\sigma_x \right)+\frac{\tilde{\eta}_a}{2}\sigma_y
\end{align}
for the optical phonons, where the $d$ coefficients were defined in Eq. (\ref{eq:ds}), and
\begin{align}
i{\cal G}=&i\frac{\eta}{2q^2}\left(q^2 \sigma_0  +2 q_x q_y \sigma_x + (q_x^2-q_y^2) \sigma_z  \right)\notag\\
+&i\frac{\tilde{\eta}_s}{q^2}\left( (q_x^2-q_y^2) \sigma_x   -2 q_x q_y\sigma_z   \right)-\frac{\tilde{\eta}_a}{2}\sigma_y
\end{align}
for the acoustic phonons. We note that $i{\cal G}$ is a Hermitian matrix.

Yet another source of ${\cal G} ({\bf q})$ is the so-called molecular Berry curvature \cite{saparov2022lattice}. An explicit calculation in App. \ref{ap:MBC} shows that, in our model, the molecular  Berry curvature remains null in the presence of a valley Zeeman splitting, insofar as the latter is not strong enough to induce a band inversion in one of the valleys. 

\subsection{Berry curvature}

We recompute the Berry curvatures for optical and acoustic phonons from  Eqs. (\ref{eq:BC}) and (\ref{eq:HDG}). Taylor-expanding first on $q$, then on $B$ and finally on $\delta a$, the leading order correction to Eq.~(\ref{eq:Berry_optical_dressed}) due to $B$ reads
\begin{equation}
\label{eq:dBerry_op}
\delta\boldsymbol{\Omega}_\pm \simeq \pm \hat{\bf z} 
\frac{g_x^o g_z^o {\cal S} a^2 \gamma_+^{3/2}{\rm sgn}(\gamma_- \delta a)}{192 \sqrt{2 M}\pi (\gamma_- \delta a)^2 |m_0|^3} (q_x^2 - q_y^2) \zeta B,
\end{equation}
for optical phonons , where we have used $M_{\rm uc}=2M$. For the acoustic phonons, the leading order correction to Eq.~(\ref{eq:Berry_acoustic_dressed}) due to $B$ reads
\begin{equation}
\label{eq:dBerry_ac}
\delta\boldsymbol{\Omega}_\pm \simeq \pm \hat{\bf z}  
\sqrt{\frac{2+\sqrt{3}}{\gamma_0 M}} \frac{g^a_x g^a_z {\cal S}}{8 \pi a  |m_0|^3} q \zeta B.
\end{equation}
Having neglected terms that are of fourth (or higher) order in the electron-phonon coupling, it turns out that both Eqs.~(\ref{eq:dBerry_op}) and (\ref{eq:dBerry_ac}) originate solely from the $\tilde{\eta}_a$ term defined in Eq. (\ref{eq:delta_eom}). 
We also note that $B$-induced corrections to the phonon Berry curvature vanish in the $q\to 0$ limit. In addition, even though those corrections arise from electron-phonon interactions, they are not directly linked to the electronic Berry curvature (unlike the Berry curvatures at $B=0$ in Sec. \ref{Sec:BC_phonon2}).

Combining Eqs. (\ref{eq:dBerry_op}) and (\ref{eq:dBerry_ac}) with Eqs. (\ref{eq:Berry_acoustic_dressed}) and  (\ref{eq:Berry_optical_dressed}), we see that 
the total Berry curvatures of optical and acoustic phonons  verify  $\boldsymbol{\Omega}_\pm(q_x, q_y, B) = -\boldsymbol{\Omega}_\pm(q_x, -q_y, -B)$ and $\boldsymbol{\Omega}_\pm(q_x, q_y, B ) = -\boldsymbol{\Omega}_\pm(-q_x, -q_y, -B)$.
These relations are consistent with the way $\boldsymbol{\Omega}_\pm({\bf q}, B)$ transforms under $C_{2x}$ and ${\cal T}$.

\subsection{Thermal Hall effect}
\label{sec:the}
Since the Berry curvature is no longer an odd function of $q_y$ when $B\neq 0$, it makes a nonvanishing contribution to the 2D thermal Hall conductivity $\kappa_{xy}$ through a term \cite{qin2012berry,sun2021phonon,zhang2016berry,zhang2019thermal}
\begin{align}
\label{eq:kxy}
\kappa_{xy}^{\rm Berry} = - \frac{k_{B}^2 T}{\hbar {\cal S}}\sum_{{\bf q},\lambda} f[\rho (\omega_{{\bf q},\lambda})] \Omega_{\lambda} ({\bf q}),
\end{align} 
where $\rho(\omega) = (e^{\hbar \omega/k_B T} - 1)^{-1}$ is the Bose-Einstein distribution
function, $\omega_{{\bf q},\lambda}$ is the phonon frequency, and $f(x)= (1 + x)(\ln [(1+x)/x)])^2 - (\ln x)^2 - 2{\rm Li}_2 (-x)$ with ${\rm Li}_2 (x)$ as the polylogarithm function of second order.
The purpose of this section is to provide an estimate of $\kappa_{xy}^{\rm Berry}$. To that end, we restore the $\hbar$ factors. 

At low temperature, the phononic thermal Hall response is dominated by long-wavelength acoustic phonons. The frequencies of the latter are obtained from the eigenvalues of the effective phonon Hamiltonian (Eq. (\ref{eq:HDG})):
\begin{equation}
\label{eq:omegaLT}
\omega_{{\bf q}, \lambda} \simeq c_\lambda q,
\end{equation}
where $c_T = a (\gamma_0/2 M)^{1/2}$ and $c_L = \sqrt{3} c_T$ are the sound velocities for the longitudinal and transverse modes. These velocities are unaffected by $B$ and by the phonon helicity.

Replacing Eqs.~(\ref{eq:omegaLT}) and (\ref{eq:dBerry_ac}) in Eq.~(\ref{eq:kxy}) and using ${\cal S} ^{-1}\sum_{\bf q} \simeq \int dq q \int_0^{2\pi} d\chi/(2\pi)^2$ with $\chi$ the polar angle of ${\bf q}$, we have
\begin{align}
\kappa_{xy}^{\rm Berry}  =  \frac{k_B^2 T}{2 \pi \hbar}   {\cal C}\int_0^{q_D} dq q^2  \ \left( f[\rho (\omega_{{\bf q}T})] - f[\rho (\omega_{{\bf q}L})] \right) ,
\end{align}
where $q_D$ is the Debye wave vector and 
\begin{equation}
{\cal C} = \sqrt{\dfrac{2+\sqrt{3}}{\gamma_0 M}} \frac{g^a_x g^a_z {\cal S}}{8 \pi a  |m_0|^3} \hbar \zeta B.
\end{equation}
In the low-temperature regime, at which only long-wavelength acoustic modes are thermally excited, the upper limit of $q$-integration can be approximated to infinity. This is justified since the function $f(x)$ vanishes when $x\rightarrow 0$. Making a change of variables $ \hbar c_\lambda q /(k_B T) \to y$ and computing a dimensionless integral, we have
\begin{equation}
\kappa_{xy}^{\rm Berry}  \simeq  0.5 \frac{k_B^5 T^4}{\hbar^3}  \frac{g^a_x g^a_z {\cal S}}{a  |m_0|^3\sqrt{\gamma_0 M}} \zeta B   \left( \frac{1}{c_T^3} - \frac{1}{c_L^3} \right).
\end{equation}
Choosing $\zeta B=5\, {\rm meV}$, $|m_0|=25\, {\rm meV}$, and other parameters as in Fig.~\ref{fig:phonondispersion} and Fig.~\ref{fig:BC}, we arrive at
\begin{equation}
\kappa_{xy}^{\rm Berry} \simeq 10^{-18}  T[{\rm K}]^4 {\rm W K}^{-1}.
\end{equation}
This thermal Hall conductance in two dimensions can be translated into the thermal Hall conductivity in three dimensions as $\kappa_{xy}^{3D}:= \kappa_{xy}^{2D}/c$, $c$ being the interlayer spacing. For BaMnSb$_2$, where $c\approx 25 \AA$~\cite{sakai2020bulk}, we have
\begin{align}
\kappa_{xy}^{\rm Berry, 3D}  &\simeq 5 \times 10^{-10} T[{\rm K}]^4 \, {\rm W m^{-1} K^{-1}}, 
\end{align}
which, at $T=40\, {\rm K}$ (this is low compared to the Debye temperature $T_D\sim \hbar c_L \sqrt{4\pi} /(a k_B)\sim500\,{\rm K}$), yields $\kappa_{xy}^{\rm Berry, 3D} \simeq 1 \, {\rm mW m^{-1} K^{-1}}$.
A thermal conductivity of this magnitude is experimentally measurable (see e.g. Ref. \cite{sharma2024phonon}), with the caveat that our estimate relies on the (in principle reasonable) choice of $5\, {\rm eV}$ for the acoustic deformation potential,  whose actual value is nevertheless not known in BaMnSb$_2$. Should we replace  $5\, {\rm eV}$ by $1\, {\rm eV}$, the estimated $\kappa_{xy}^{\rm Berry}$ would drop by a factor $25$.

The low-temperature $T^4$ behavior of $\kappa_{xy}^{\rm Berry}$ coming from Dirac fermions is different from the $T^3$ dependence expected in magnetic solids \cite{qin2012berry} and the $T^2$ dependence anticipated in nonmagnetic insulators from a different physical origin (the molecular Berry curvature induced by a magnetic field) \cite{saito2019berry}. Incidentally, the numerical value of $\kappa_{xy}^{\rm Berry}$ at $T\sim 10 K$ is orders of magnitude larger in our theory than it is in Ref. \cite{saito2019berry}.

\section{Summary and Conclusions}
To summarize, taking BaMnSb$_2$ as a prototypical ${\cal T}$-symmetric Dirac insulator, we have shown that phonons acquire a Berry curvature by virtue of their coupling to Dirac fermions. The phonon Berry curvature being proportional to the electronic valley Chern number, we have characterized it as being ``inherited". This inherited Berry curvature is fundamentally different from the so-called molecular Berry curvature, which is an electronic Berry curvature resulting from ionic displacements in materials lacking ${\cal T}$-symmetry. 
While our calculation has concentrated on a low-energy effective model for BaMnSb$_2$, a similar effect could be realized in other ${\cal T}$-symmetric two-dimensional and three-dimensional materials with nontrivial electronic topology. 

For completeness, we have considered the effect of a ${\cal T}$-breaking perturbation. 
We have shown that an electronic valley Zeeman splitting modifies the phonon self-energy and the phonon Berry curvature through the electron-phonon interaction, thereby causing an intrinsic thermal Hall effect of phonons, even when the molecular Berry curvature is zero. This finding may have implications in the general theoretical understanding of the phonon Hall effect.

A legitimate critique for the present work concerns our definition of the ``bare'' phonons in Sec. \ref{sec:bare_phonons}. To describe the lattice vibrations in the absence of the coupling to Dirac fermions, we have adopted a phenomenological force constant matrix that is symmetric under the interchange of the cartesian indices (Eq. (\ref{eq:force_constant_symmetry})). We have shown that this symmetry results in phonons that do not have a Berry curvature. While the type of phenomenological model we have employed is commonplace, we have no proof that it is satisfactory to describe the phonons of BaMnSb$_2$ in the absence of Dirac fermions. In other words, while we can attest that Dirac fermions in BaMnSb$_2$ generate a phonon helicity and a phonon Berry curvature, we cannot assert that the latter originate {\rm exclusively} from the Dirac fermions. This issue can be settled by first-principles calculations of the phonon dispersion and polarizations.

Another limitation of our analytical theory is that it is restricted to long-wavelength phonons. It would be interesting to generalize it for arbitrary phonon wave vectors, notably in order to find out whether phonon bands in BaMnSb$_2$ could acquire a nonzero Chern number in the presence of an electronic valley Zeeman splitting.

Yet another avenue for future work consists of exploring the possibility in which topologically trivial bare electrons would acquire a Berry curvature through their coupling to topologically nontrivial phonons. 
This effect would be converse to the one discussed in the present paper. 
While phonon-induced electronic topological transitions were predicted earlier \cite{garate2013phonon}, we are not aware of instances in which the electronic Berry curvature is known to emerge from the electron-phonon interaction.

\section*{Acknowledgments}
The authors acknowledge funding from the Canada First Research Excellence Fund and the Natural Sciences and Engineering Research Council of Canada (Grants No. RGPIN-2018-05385 and RGPIN-2024-05210). S. G. acknowledges financial support from the Anusandhan National Research Foundation (ANRF) under Grant No. ANRF/ECRG/2024/001412/PMS and Department of Science and Technology (DST) under DST-FIST Project No. SR/FST/PSI/2017/5(C). We thank anonymous referees for insightful comments. I.G. is grateful for the hospitality of the Donostia International Physics Center in the final stages of the present work. The authors benefited from their affiliation to the RQMP, a strategic research network funded by the Fonds de Recherche du Québec~\cite{rqmp}.


\begin{widetext}

\appendix

\section{Dynamical matrix for bare in-plane phonons}
\label{ap:dynamical}

For reference, the explicit expressions for the non-null elements of the dynamical matrix in Eq.~(\ref{eq:dyn}) are as follows:
\begin{align}
\label{eq:list}
D_{xx}(1,1) &= \frac{2}{M}\left[ \gamma_2 \left(\frac{\delta a + 1}{A} \right)^2 + \gamma_1 \left(\frac{\delta a-1}{B} \right)^2 + \gamma_0 (1 - \cos q_x a) \right] \notag\\
D_{yy}(1,1) &= \frac{2}{M}\left[ \gamma_2 \left(\frac{1}{A} \right)^2 + \gamma_1 \left(\frac{1}{B} \right)^2 + \gamma_0 (1 - \cos{q_y a}) \right]\notag\\
D_{xx}(1,2) &= \frac{2}{M} \cos \frac{q_y a}{2} \left[ - \gamma_2 \left( \frac{\delta a + 1}{A} \right)^2 e^{i \frac{q_x a}{2}(\delta a + 1)} - \gamma_1 \left( \frac{\delta a - 1}{B} \right)^2 e^{i \frac{q_x a}{2}(\delta a - 1)} \right]\notag\\
D_{yy}(1,2) &= \frac{2}{M} \cos \frac{q_y a}{2} \left[ - \gamma_2 \left( \frac{1}{A} \right)^2 e^{i \frac{q_x a}{2}(\delta a + 1)} - \gamma_1 \left( \frac{1}{B} \right)^2 e^{i \frac{q_x a}{2}(\delta a - 1)} \right]\notag\\
D_{xy}(1,2) &= i \frac{2}{M} \sin \frac{q_y a}{2} \left[ - \gamma_2 \left( \frac{\delta a + 1}{A^2} \right) e^{i \frac{q_x a}{2}(\delta a + 1)} - \gamma_1 \left( \frac{\delta a - 1}{B^2} \right) e^{i \frac{q_x a}{2}(\delta a - 1)} \right]\notag\\
D_{xx}(2,2) &= D_{xx}(1,1)\notag\\
D_{yy}(2,2) &= D_{yy}(1,1)\notag\\
D_{yx}(1,2) &= D_{xy}(1,2),
\end{align}
with $A=\sqrt{1 + (1+\delta a)^2}$ and $B=\sqrt{1 + (1-\delta a)^2}$. 
These results were first obtained in Ref. \cite{hu2021phonon} using a force-constant model with only bond-stretching contributions to the lattice potential energy. Only force-constants between first and second-nearest neighbors were kept, and the unit cell was approximated as a square (of unit length in the equations above).

In the present work, we have considered two more general models: 
(1) a model that extends the one of Ref. \cite{hu2021phonon} by keeping up to third-neighbor force constants in a rectangular (rather than square)  lattice;
(2) an ``axially symmetric'' model, which augments the model of Ref. \cite{hu2021phonon} by incorporating bond-bending contributions to the force constants \cite{lehman1962axially, pintschovius1983lattice}:
 \begin{align}
\Phi_{\alpha\beta}(l \kappa,l^\prime \kappa^\prime)=\gamma(l\kappa,l^\prime \kappa^\prime)\hat{\bf e}_{\alpha}\hat{\bf e}_{\beta},
+\tilde{\gamma}(l\kappa,l^\prime,\kappa^\prime)(\delta_{\alpha \beta}-\hat{\bf e}_{\alpha} \hat{\bf e}_{\beta}),
\end{align}
 $\tilde{\gamma}(l\kappa,l^\prime,\kappa^\prime)$ being the transverse spring constants.

The generalization \# 1 modifies the elements in Eq.~(\ref{eq:list}) and gives nonzero values to elements that were zero in Eq.~(\ref{eq:list}) (such as $D_{xy}(1,1)$).
The generalization \#2 also modifies the nonzero elements in Eq.~(\ref{eq:list}). 
In both of these generalized models, we have obtained the effective $2\times 2$ dynamical matrices for long wavelength acoustic and optical phonons, thereafter verifying that those phonons have a vanishing helicity and Berry curvature.
The underlying reason, as proven in the main text, is that the force constant models we have adopted for bare phonons obey Eq. (\ref{eq:force_constant_symmetry}).

\section{Effective phonon action}
\label{ap:action}

As a consequence of the electron-phonon interaction, the phonon effective action for optical and acoustic modes acquires a Chern-Simons term. Following the first prediction of Ref.~\cite{hu2021phonon}, it was confirmed in Ref.~\cite{parlak2023detection} for optical phonons using functional integration. Here, we carry out the same for acoustic modes as well and lay down the key steps here. 

The partition function of the coupled electron-phonon system can be written as
\begin{equation}
Z=\int_{c, \overline{c}, u} e^{-S[u, c, \overline{c}]},
\end{equation}
 where $\int_x$ stands for a path integral over the field $x$,  $u$ is the phonon field, $c$ and $\overline{c}$ are fermion fields, and 
 \begin{equation}
 S[v,c,\overline{c}]= S_e[c,\overline{c}]+S_{\rm ph}[u]+S_{\rm e-ph}[u, c, \overline{c}]
 \end{equation}
 is the action in imaginary time with
 \begin{align}
 S_e[c,\overline{c}] &=\sum_{{\bf k}, k_n} \overline{c}_{\bf k}(k_n) \left(-i k_n+h({\bf k}) \right)c_{\bf k}(k_n)\notag\\
 S_{\rm ph}[u]&=\frac{M_{\rm uc}}{2} \sum_{{\bf q}, q_m, \lambda} (q_m^2+\omega_{\lambda,{\bf q}}^2) u_{-{\bf q}, \lambda}(-q_m) u_{{\bf q},\lambda} (q_m)\notag\\
 S_{\rm e-ph}[u, c, \overline{c}] &=\sqrt{k_B T}\sum_{{\bf k}, {\bf q}, \lambda}\sum_{k_n, q_m} \overline{c}_{\bf k}(k_n) g^\lambda({\bf k}, {\bf q}) c_{{\bf k}-{\bf q}}(k_n-q_m) u_{{\bf q}, \lambda}(q_m).
 \end{align}
 Here, $k_n = (2n + 1) k_B T$ ($n\in\mathbb{Z}$) and $ q_m = 2m k_B T$  ($m\in\mathbb{Z}$) refer to fermionic and bosonic Matsubara frequencies, respectively, with $T$ the temperature. Also, $c$ and $\overline{c}$ are 4-component spinor fields, due to valley and orbital degrees of freedom. Likewise, $g^\lambda$ is a $4\times 4$ matrix for the coupling between electrons and the phonon mode $\lambda$.
In the preceding expressions, we have adopted the Fourier transform convention
\begin{align}
f(\tau) &=\sqrt{k_B T}\sum_{k_n} e^{-i k_n \tau} f(k_n)\notag\\
f(k_n) &=\sqrt{k_B T}\int_0^{1/(k_B T)} e^{i k_n \tau} f(\tau).
\end{align}
Thus, the dimensions of $u_{{\bf q},\lambda} (q_m)$ are ${\rm length}\times\sqrt{\rm time}$. 
We also note that Coulomb interaction between electrons has been neglected above; their effect amounts to renormalization of the electron-phonon vertices (cf. Ref. \cite{parlak2023detection}), which is not essential for the present discussion.
 
Next, we integrate out the electron fields in order to get
\begin{equation}
Z=\int_{u} e^{-S_{\rm eff}[u]},
\end{equation}
with
\begin{align}
S_{\rm eff}&=\frac{M_{\rm uc}}{2} \sum_{{\bf q}, q_m} \left(u^*_{{\bf q}, 1} (q_m), u^*_{{\bf q}, 2}(q_m)\right) \left(\begin{array}{cc}  q_m^2 + \omega_{{\bf q}, 1}^2 + \Sigma_{11}({\bf q},q_m) & \Sigma_{12}({\bf q},q_m) \\ \Sigma_{21}({\bf q},q_m) &  q_m^2 + \omega_{{\bf q}, 2}^2+\Sigma_{22}({\bf q},q_m) \end{array}\right) \left(\begin{array}{c}u_{{\bf q}, 1}(q_m) \\ u_{{\bf q}, 2}(q_m) \end{array}\right),
\end{align}
with the notation of this equation discussed in the main text. For optical phonons ($1\equiv A$, $2\equiv B$), the phonon self-energies are given as
\begin{align}
\label{eq:self_op}
\Sigma_{AA}({\bf q},q_m)&= \frac{{g^o_0}^2}{M_{\rm uc}}  \Pi_{00}^{(0)}({\bf q},q_m) + \frac{{g^o_y}^2}{M_{\rm uc}}  \Pi_{yy}^{(0)}({\bf q},q_m) + \frac{{g^o_z}^2}{M_{\rm uc}}  \Pi_{zz}^{(0)}({\bf q},q_m) \nonumber\\
 &+ \frac{1}{M_{\rm uc}} \left[g^o_y g^o_0  \Pi_{0y}^{(z)}({\bf q},q_m) +g^o_z g^o_0  \Pi_{0z}^{(0)}({\bf q},q_m) + g^o_y g^o_z \Pi_{zy}^{(z)}({\bf q},q_m) + ({\bf q},q_m \leftrightarrow -{\bf q},-q_m) \right]\nonumber\\
  &=\Sigma_{AA}(-{\bf q},-q_m),\nonumber\\
  \Sigma_{BB}({\bf q},q_m)&=  \frac{{g^o_x}^2}{M_{\rm uc}}  \Pi_{xx}^{(0)}({\bf q},q_m) \ = \Sigma_{BB}(-{\bf q},-q_m),\nonumber\\
  \Sigma_{AB}({\bf q},q_m)&= \frac{1}{M_{\rm uc}} \left( g^o_x g^o_0  \Pi_{0x}^{(0)}(-{\bf q},-q_m) + g^o_x g^o_y \Pi_{yx}^{(z)}(-{\bf q},-q_m)+ g^o_x g^o_z \Pi_{zx}^{(0)}(-{\bf q},-q_m)\right) =\Sigma_{BA}(-{\bf q},-q_m),
 \end{align}
whereas the same for acoustic modes [$1\equiv T$ (for ``transverse''), $2\equiv L$] reads 
\begin{align}
\label{eq:self_ap}
  \Sigma_{LL}({\bf q},q_m)&=  \frac{q^2}{M_{\rm uc}} {g^a_x}^2  \Pi_{xx}^{(0)}({\bf q},q_m) \ = \Sigma_{LL}(-{\bf q},-q_m),\nonumber\\
  \Sigma_{TT}({\bf q},q_m)&= \frac{q^2}{M_{\rm uc}} \left[{g^a_0}^2  \Pi_{00}^{(0)}({\bf q},q_m) + {g^a_y}^2  \Pi_{yy}^{(0)}({\bf q},q_m) + {g^a_z}^2  \Pi_{zz}^{(0)}({\bf q},q_m)\right] \nonumber\\
 &+ \frac{q^2}{M_{\rm uc}} \left[g^a_y g^a_0  \Pi_{0y}^{(z)}({\bf q},q_m) +g^a_z g^a_0  \Pi_{0z}^{(0)}({\bf q},q_m) + g^a_y g^o_z \Pi_{zy}^{(z)}({\bf q},q_m) + ({\bf q},q_m \leftrightarrow -{\bf q},-q_m) \right]\nonumber\\
  &=\Sigma_{TT}(-{\bf q},-q_m),\nonumber\\
  \Sigma_{TL}({\bf q},q_m)&= \frac{q^2}{M_{\rm uc}} \left[ g^a_x g^a_0  \Pi_{0x}^{(0)}(-{\bf q},-q_m) + g^a_x g^a_y \Pi_{yx}^{(z)}(-{\bf q},-q_m)+ g^a_x g^a_z \Pi_{zx}^{(0)}(-{\bf q},-q_m)\right]\ =\Sigma_{LT}(-{\bf q},-q_m).
 \end{align}
Here, $\Pi^{(0)}_{ij} ({\bf q}, q_m )$ and $\Pi^{(z)}_{ij} ({\bf q}, q_m )$
are the elements of linear response tensor for electrons given as
\begin{align}
\label{eq:resp_funcs}
\Pi^{(0)}_{ij} ({\bf q}, q_m ) &=k_B T \sum_{s,{\bf k}, k_n} 
{\rm Tr} [\tau_i G^s_{0,e} ({\bf k}, k_n)\tau_j G^s_{0,e} ({\bf k}+{\bf q} ,  k_n + q_m )]\\
\Pi^{(z)}_{ij} ({\bf q}, q_m ) &= k_B T \sum_{s,{\bf k}, k_n} 
s \ {\rm Tr} [\tau_i G^s_{0,e} ({\bf k}, k_n)\tau_j G^s_{0,e} ({\bf k}+{\bf q} ,  k_n + q_m )],
\end{align}
where $\tau_i$ are Pauli matrices, $s$ labels the two valleys and $G^s_{0,e} ({\bf k},k_n) = (i k_n - h^s_e ({\bf k}))^{-1}$ is the bare Matsubara Green’s function of Dirac electrons at valley $s$. Also, we have assumed the Fermi energy inside the gap of insulator ($\mu = 0$).

The Matsubara sums in Eq.~(\ref{eq:resp_funcs}) are easily computed. Afterwards, the analytically continued response functions $\Pi^{(0)}_{ij} ({\bf q}, \omega )$ and $\Pi^{(z)}_{ij} ({\bf q}, \omega)$ are obtained via $q_m \longrightarrow - i \omega + 0^+$.
These analytically continued functions have in general both real and imaginary components, which can be extracted by using the Plemelj-Sokhotski theorem. 
In the absence of disorder and at low temperature, only the principal value of the ${\bf k}$-integral contributes when the phonon frequency is smaller than the energy gap of the insulator. 
The physical reason for this is that low-frequency phonons cannot excite real particle-hole pair across the energy gap of the insulator.
This is the physically relevant regime for long-wavelength acoustic phonons. 
It can also be applicable to long-wavelength optical phonons (in BaMnSb$_2$, the direct energy gap at the $K_\pm$ valleys is about 50 meV \cite{liu2021spin} and the typical optical phonon frequency at $q\to 0$ is less than 30 meV \cite{chen2024thermoelectric}).

The first element we look at is the charge response $\Pi^{(0)}_{00}$. In the low-temperature limit, it reads
\begin{align}
\Pi_{00}^{(0)}({\bf q},  q_m) &= - \frac{{\cal S}}{(2\pi)^2} \sum_{s=\pm 1}\int d^2k \frac{(E_k + E_{k^{\prime}}) (E_k E_{k^{\prime}} - m^2_0 - v_0^2 {\bf k}\cdot {\bf k}^{\prime})}{E_k E_{k^{\prime}} [(E_k + E_{k^{\prime}})^2 + q_m^2]},
\label{Eq:Pi00}
\end{align}
where ${\cal S}$ is the sample area, ${\bf k^{\prime}} = {\bf k} + {\bf q}$, $E_k = \sqrt{v_0^2 k^2 + m_0^2}$, $k^2=k_x^2 + k_y^2$. Taking analytical continuation and assuming the phonon frequency $\omega$ to be much smaller then the insulating gap ($2 |m_0|$), we obtain 
$
\Pi_{00}^{(0)}({\bf q}, \omega) = - q^2 {\cal S}/(6 \pi |m_0|)
$.

Now, we look at the diagonal pseudospin response elements $\Pi_{ii}$ ($i=x, y, z$). Owing to the the presence of the infinite sea of negative-energy states in our continuum model, $\Pi_{xx}$ and $\Pi_{zz}$ contains a unphysical gauge-violating term and should be regularized by taking the $\mu = 0$ ground state as the reference system ~\cite{ghosh2019charge,polini2009drude,sabio2008fsum}. This amounts to including the anomalous commutator term $\langle [\hat{\tau}_i ({\bf q}), \hat{\rho}(-{\bf q})] \rangle = q \Lambda/(2\pi)  $ in $\Pi_{xx}$ and $\Pi_{zz}$, with $\hat{\tau}_i ({\bf q})= \sum_{s,\alpha,\beta, {\bf k}}c^\dagger_{s,\alpha,{\bf k}} \tau^i_{\alpha \beta} c_{s,\beta,{\bf k} - {\bf q}}$ and $\hat{\rho} ({\bf q})= \sum_{s,\alpha, {\bf k}}c^\dagger_{s,\alpha,{\bf k}} c_{s,\alpha,{\bf k} - {\bf q}}$ being the electron's pseudospin and density operator, respectively and $\Lambda$ is the ultraviolet cutoff for the momentum sum.

Considering such regularization, we obtain
$
\Pi_{xx}^{(0)}({\bf q}, \omega)/{\cal S} = - q_y^2/(6 \pi |m_0|)
$,
and
$
\Pi_{zz}^{(0)}({\bf q}, \omega)/{\cal S} = - q_x^2/(6 \pi |m_0|)
$ in the low-frequency limit. However, for $\Pi^{(0)}_{yy}$, there is no anomalous commutator term, and we obtain
$
\Pi^{(0)}_{yy} ({\bf q}, \omega) /{\cal S} = - \Lambda/(\pi v_0)
$.

We now carry out similar calculation for off-diagonal elements when the phonon frequency is smaller than the energy gap. For example,  
\begin{align}
\Pi^{(0)}_{0x} ({\bf q},\omega) /{\cal S}&= i  v_0 \frac{1}{(2\pi)^2} \sum_{s=\pm 1}\int d^2k  \frac{(E_k + E_{k^{\prime}})m_0 q_y + i s q_x \omega E_k}{E_k E_{k^{\prime}} [(E_k + E_{k^{\prime}})^2 - \omega^2]},\nonumber \\
&=\frac{i v_0}{(2\pi)^2} \int d^2k \frac{2(E_k + E_{k^{\prime}})m_0 q_y }{E_k E_{k^{\prime}} [(E_k + E_{k^{\prime}})^2 - \omega^2]},
\label{Eq:Pi01}
\end{align}
which, for $ \omega \ll 2|m_0|$ and long-wavelength limit, simplifies to
\begin{equation}
\Pi^{(0)}_{0x} ({\bf q},\omega)/{\cal S} = \frac{i q_y}{2\pi v_0} {\rm sgn} (m_0).
\end{equation}
Carrying out similar calculation and taking terms up to first-order in $q$, we obtain
\begin{align}
\Pi^{(z)}_{0y} ({\bf q},\omega) &= 0,\notag \\
\Pi^{(0)}_{0z} ({\bf q},\omega)/{\cal S} &=  - \frac{i q_x}{2\pi v_0} {\rm sgn} (m_0),\notag \\
\Pi^{(z)}_{xy} ({\bf q},\omega) &=  0, \notag \\
\Pi^{(0)}_{xz} ({\bf q},\omega) &=  0,\notag \\
\Pi^{(z)}_{yz} ({\bf q},\omega) &=  0.
\end{align}
Other off-diagonal responses can be obtained by using hermiticity $\Pi^{(0/z)}_{ij} ({\bf q},\omega) = \left( \Pi^{(0/z)}_{ji} ({\bf q},\omega) \right)^*$. 

Therefore, in the long-wavelength, low-frequency limit, considering terms in $\Pi_{ij}^{0/z}$ that are up to first-order in $q$, we have
\begin{align}
\Sigma_{AA}({\bf q},\omega) &=  - \frac{{g^o_2}^2 {\cal S}}{M_{\rm uc}} \frac{\Lambda}{2\pi v_0}\nonumber \\
\Sigma_{BB}({\bf q},\omega) &= 0 \nonumber \\
\Sigma_{AB}({\bf q},\omega) &= -\Sigma_{BA}({\bf q},\omega) =-\frac{g^o_0 g^o_x {\cal S}}{M_{\rm uc}} \frac{i q_y}{2\pi v_0} {\rm sgn} (m_0).
\end{align}
for optical phonons, and
\begin{align}
\Sigma_{LL}({\bf q},\omega) &= 0 \nonumber \\
\Sigma_{TT}({\bf q},\omega) &=  - \frac{{g^a_2}^2 q^2  {\cal S}}{M_{\rm uc}} \frac{\Lambda}{2\pi v_0}\nonumber \\
\Sigma_{LT}({\bf q},\omega) &=-\Sigma_{TL}({\bf q},\omega) = \frac{g^a_0 g^a_x q^2 {\cal S}}{M_{\rm uc}}  \frac{i q_y}{2\pi v_0} {\rm sgn} (m_0). 
\end{align}
for acoustic phonons.
Thus, in this regime, the diagonal elements of the phonon self-energy are purely real whereas the off-diagonal elements are purely imaginary.

\section{Influence of a Zeeman field on the phonon self-energy}
\label{sec:delta_self}

In this Appendix, we show how Eqs. (\ref{eq:self_op}) and (\ref{eq:self_ap}) are modified in the presence of a valley Zeeman splitting $2\zeta B$ (i.e., when the masses of the Dirac fermions at $K$ and $K'$ are $m_0+\zeta B$ and $-m_0+\zeta B$, respectively. Here, $B$ is an external magnetic perturbation that breaks time-reversal symmetry. 
We will assume that $B$ does not modify the electronic structure of the massive Dirac fermions, aside from the valley Zeeman splitting. 
We will furthermore suppose that $|\zeta B|\ll |m_0|$.

Then, the changes of the zero-temperature phonon self-energy to first order in $B$ and in the long-wavelength approximation originate from 
\begin{align}
\delta\Sigma_{AA}({\bf q}, q_m) &\simeq \frac{g_z^o g_0^o {\cal S}}{M_{\rm uc}} \frac{i q_y q_m}{6\pi v_0 m_0^2} \zeta B\notag\\
\delta\Sigma_{AB}({\bf q}, q_m)&\simeq \frac{g_x^o g_0^o  {\cal S}}{M_{\rm uc}} \frac{i q_x q_m}{6\pi v_0 m_0^2} \zeta B +\frac{g_x^o g_z^o  {\cal S}}{M_{\rm uc}} \frac{q^2 q_m}{12\pi |m_0|^3} \zeta B \notag\\
\delta\Sigma_{TT}({\bf q}, q_m) &\simeq \frac{g_z^a g_0^a q^2  {\cal S}}{M_{\rm uc}} \frac{i q_y q_m}{6\pi v_0 m_0^2} \zeta B\notag\\
\delta\Sigma_{TL}({\bf q}, q_m)&\simeq \frac{g_x^a g_0^a q^2  {\cal S}}{M_{\rm uc}} \frac{i q_x q_m}{6\pi v_0 m_0^2} \zeta B +\frac{g_x^a g_z^a q^2  {\cal S}}{M_{\rm uc}} \frac{q^2 q_m}{12\pi |m_0|^3} \zeta B.
\end{align} 
These expressions follow directly from Eqs. (\ref{eq:self_op}) and (\ref{eq:self_ap}), upon incorporating the valley Zeeman splitting.
In the long-wavelength approximation, the first terms on the right hand sides of $\delta\Sigma_{AB}$ and $\delta\Sigma_{TL}$ are dominant over the second terms. Nevertheless, the latter make the leading contribution to the phonon Berry curvature. We therefore keep them here and in the main text.

\section{Molecular Berry curvature for in-plane phonons in BaMn$_2$Sb$_2$}
\label{ap:MBC}
In the Born-Oppenheimer approximation, the electronic ground state can accumulate nontrivial geometric phase when the ionic configuration evolves. This electronic Berry phase with respect to the ions' displacement is termed as molecular Berry phase and the corresponding Berry curvature is called the molecular Berry curvature~\cite{mead1992geometric,saparov2022lattice}. Here, we calculate the molecular Berry curvature (MBC) for the in-plane phonons of BaMn$_2$Sb$_2$ and show that it vanishes as long as the two valleys have energy gaps of opposite signs (even when ${\cal T}$ symmetry is broken by an exchange field through the valley Zeeman splitting). 

The MBC tensor elements at zero temperature can be written as \cite{saparov2022lattice}
\begin{align}
G^{\rm MBC} (\bf{q})=  \begin{pmatrix}
G_{xx}(1,1;{\bf q}) & G_{xy}(1,1;{\bf q}) & G_{xx}(1,2;{\bf q}) & G_{xy}(1,2;{\bf q})  \\[.5em]
G_{yx}(1,1;{\bf q}) & G_{yy}(1,1;{\bf q}) & G_{yx}(1,2;{\bf q}) & G_{yy}(1,2;{\bf q})  \\[.5em]
G_{xx}(2,1;{\bf q}) & G_{xy}(2,1;{\bf q}) & G_{xx}(2,2;{\bf q}) & G_{xy}(2,2;{\bf q})  \\[.5em]
G_{yx}(2,1;{\bf q}) & G_{yy}(2,1;{\bf q}) & G_{yx}(2,2;{\bf q}) & G_{yy}(2,2;{\bf q})  
\end{pmatrix},
\end{align}
where
\begin{align}
\label{eq:G}
G_{\alpha\beta}(\kappa,\kappa';{\bf q})=  \frac{i}{N}  \sum_{{\bf k}, s} \sum_{\substack{\epsilon_n<0\\\epsilon_{n'}>0}} &\left[\frac{\langle \phi_{n, s} ({\bf k})| {\cal M}_{\alpha}^\kappa ({\bf q}, s)| \phi_{n', s} ({\bf k} + {\bf q}) \rangle \langle \phi_{n', s} ({\bf k} + {\bf q}) | {\cal M}_{\beta}^{\kappa'} (-{\bf q}, s) | \phi_{n, s} ({\bf k}) \rangle}{\left(\epsilon_{n, s} ({\bf k}) - \epsilon_{n', s} ({\bf k}+{\bf q})\right)^2}\right.\notag\\
&-\left. \frac{\langle \phi_{n, s} ({\bf k}+{\bf q})| {\cal M}_{\beta}^{\kappa'} (-{\bf q}, s)| \phi_{n', s} ({\bf k}) \rangle \langle \phi_{n', s} ({\bf k}) | {\cal M}_{\alpha}^{\kappa} ({\bf q}, s) | \phi_{n, s} ({\bf k}+{\bf q}) \rangle}{\left(\epsilon_{n, s} ({\bf k}+{\bf q}) - \epsilon_{n', s} ({\bf k})\right)^2}\right], 
\end{align}
${\bf q}$ is the phonon wave vector, $N$ is the number of unit cells in the crystal,
\begin{equation}
\label{eq:M}
{\cal M}_{\alpha}^\kappa (-{\bf q}, s) =  \sqrt{N}\sum_\lambda g^{\lambda, s}({\bf q}) \frac{\partial u_{{\bf q} \lambda}}{\partial w_{{\bf q}, \kappa,\alpha}}
\end{equation}
is the electron-phonon interaction vertex for an electron in valley $s$ when an atom $\kappa$ undergoes a displacement $w_{{\bf q}, \kappa,\alpha}$ along axis $\alpha$, $|\phi_{n, s} ({\bf k})\rangle$ is the eigenstate of the free electron Hamiltonian (Eq.~(\ref{eq:hek})) in band $n$ at wave vector ${\bf k}$ and valley $s$, $\epsilon_{n, s}({\bf k})$ is the corresponding energy eigenvalue, $\sum_{\bf k} \equiv {\cal S} \int d^2k/(2\pi)^2$ and ${\cal S}$ is the surface area of the sample. 

The derivative in Eq. (\ref{eq:M}) can be calculated from the relation
\cite{rinkel2017signatures}
\begin{equation}
{\bf u}_{{\bf q} \lambda} =\frac{1}{2 M}\sum_\kappa M_\kappa {\bf w}_{{\bf q}, \kappa} \cdot {\bf p}^*_{{\bf q} \lambda, \kappa} e^{-i\phi_\lambda},
\end{equation}
where ${\bf p}_{{\bf q}\lambda, \kappa}$ is the polarization vector denoting the displacement of atom $\kappa$ in normal mode $({\bf q},\lambda)$,
$\phi_\lambda$ is a (for now arbitrary) phase factor that originates from the fact that the polarization vectors ${\bf p}_{{\bf q}\lambda, \eta}$ of a given mode $({\bf q}, \lambda)$ are defined modulo an $\kappa$-independent phase factor, $M_\kappa$ is the mass of atom $\kappa$, $2 M$ is the atomic mass in the unit cell of two Sb atoms,
\begin{equation}
{\bf w}_{{\bf q}, \kappa} = \frac{1}{N} \sum_{\bf l} e^{-i {\bf q}\cdot{\bf l}} {\bf w}_{{\bf l}, \kappa},
\end{equation}
and ${\bf w}_{{\bf l}, \kappa}$ is the displacement of atom $\kappa$ in a unit cell of position ${\bf l}$. The hermiticity conditions ${\bf w}_{{\bf l}, \kappa}^\dagger={\bf w}_{{\bf l}, \kappa}$ and $u^\dagger_{{\bf q}\lambda} = u_{-{\bf q}\lambda}$ impose
\begin{equation}
\label{eq:phi_lambda}
e^{i\phi_\lambda} {\bf p}_{{\bf q}\lambda, \kappa} = e^{-i\phi_\lambda} {\bf p}^*_{-{\bf q}\lambda, \kappa}.
\end{equation}
In the main text, we have taken the polarization vector of the optical phonons to be real and even in ${\bf q}$, while those of acoustic phonons were real and odd in ${\bf q}$. 
Consequently, Eq.~(\ref{eq:phi_lambda})  implies that $\phi_\lambda=0$ for optical phonons and $\phi_\lambda=\pi/2$ for acoustic phonons (we could have equally well taken $\phi_\lambda=-\pi/2$).
 
With the preceding caveats, it follows that
\begin{equation}
{\cal M}_{\alpha}^\kappa (-{\bf q}, s) = \sqrt{N}\frac{1}{2}\sum_\lambda g^{\lambda, s}({\bf q}) p^*_{{\bf q} \lambda, \kappa, \alpha} e^{-i \phi_\lambda},
\end{equation}
with ${\cal M}_\alpha^\kappa ({\bf q}, s) ={\cal M}_{\alpha}^\kappa (-{\bf q}, s) ^\dagger$.

For long-wavelength phonons, we have
\begin{align}
\label{eq:Mnueta}
{\cal M}_{\alpha}^\kappa ({\bf q}, s)/\sqrt{N} &=\frac{1}{2}(\delta_{\kappa, 1}-\delta_{\kappa, 2})  \left(g^{A_1, s} \delta_{\alpha, x} + g^{B_1, s} \delta_{\alpha, y}\right)\notag\\
&+\frac{i}{2}(\delta_{\kappa, 1}+\delta_{\kappa, 2})\left[\left(g^{LA, s} \cos\chi+g^{TA, s} \sin\chi\right) \delta_{\alpha, x}+ \left(g^{LA, s} \sin\chi-g^{TA, s} \cos\chi\right) \delta_{\alpha, y}\right],
\end{align}
where $\delta_{i j}$ is Kronecker's delta, $\chi$ is the angle between $\hat{\bf q}$ and $\hat{\bf x}$, and the $i$ in the second line comes from the fact that $\phi_\lambda=\pi/2$ for acoustic phonons. 
The first and second lines in Eq. (\ref{eq:Mnueta}) originate from optical and acoustic phonons, respectively.
 The couplings $g^{A_1,s}$ and $g^{B_1, s}$ are defined in Eq. (\ref{eq:adef0}), while
the couplings $g^{LA, s}$ and $g^{TA, s}$ are defined in Eqs. (\ref{eq:adef}) and (\ref{eq:adef2}).

Substituting Eq. (\ref{eq:Mnueta}) in Eq. (\ref{eq:G}), we compute the MBC explicitly. 
When ${\cal T}$-symmetry is present, Eq.~(\ref{eq:G}) vanishes identically due to perfect cancellation between the contributions from the two electronic valleys.
Specifically, if we write
\begin{equation}
G_{\alpha \beta}  (\kappa,\kappa';{\bf q}) \equiv \sum_s G_{\alpha \beta}  (\kappa,\kappa';{\bf q}, s),
\end{equation}
then we find $G_{\alpha \beta}  (\kappa,\kappa';{\bf q}, +) = -G_{\alpha \beta}  (\kappa,\kappa';{\bf q}, -)$ in the presence of ${\cal T}$-symmetry.
It is nonetheless instructive to list the matrix elements of $G_{\alpha \beta}  (\kappa,\kappa';{\bf q}, s)$ to leading order in $q$:
\begin{align}
\label{eq:calG}
G_{xx} (1,1;{\bf q}, s)&=-s \frac{i}{8\pi} \frac{{\rm sign}(m_0)}{v_0^2} q g_z^o g_x^a {\cal S} \cos\chi\notag\\
G_{yy} (1,1;{\bf q}, s) &= -s \frac{i}{8\pi} \frac{{\rm sign}(m_0)}{v_0^2} q g_x^o g_z^a {\cal S} \cos\chi\notag\\
G_{xx} (2,2; {\bf q}, s) &=-G_{xx} (1,1; {\bf q}, s) \notag\\
G_{yy} (2,2; {\bf q}, s) &=-G_{yy} (1,1; {\bf q}, s) \notag\\
G_{xy} (1,1; {\bf q}, s) &=s \frac{1}{16\pi}\frac{{\rm sign}(m_0)}{v_0^2} \left(g_x^o g_z^o + i q \left(g_x^o g_z^a- g_x^a g_z^o\right) \sin\chi\right){\cal S}\notag\\
G_{yx} (1,1; {\bf q}, s) &=-s \frac{1}{16\pi}\frac{{\rm sign}(m_0)}{v_0^2} \left(g_x^o g_z^o - i q \left(g_x^o g_z^a- g_x^a g_z^o\right) \sin\chi\right){\cal S}\notag\\
G_{xx} (1,2; {\bf q}, s) &= G_{xx} (2,1; {\bf q}, s) = 0 \notag\\
G_{xy} (1,2; {\bf q}, s) &=s \frac{1}{16\pi}\frac{{\rm sign}(m_0)}{v_0^2} \left(-g_x^o g_z^o - i q \left(g_x^o g_z^a + g_x^a g_z^o\right) \sin\chi\right){\cal S}\notag\\
G_{yx} (2,1; {\bf q}, s) &=-s \frac{1}{16\pi}\frac{{\rm sign}(m_0)}{v_0^2} \left(-g_x^o g_z^o + i q \left(g_x^o g_z^a+ g_x^a g_z^o\right) \sin\chi\right){\cal S}\notag\\
G_{yx} (1,2; {\bf q}, s) & = - G_{xy} (1,2; {\bf q}, s)\notag\\
G_{xy} (2,1; {\bf q}, s) & =- G_{yx} (2,1; {\bf q}, s)\notag\\
G_{yy} (1,2; {\bf q}, s) & =G_{yy} (2,1; {\bf q}, s) = 0\notag\\
G_{xy} (2,2; {\bf q}, s) &= -G_{yx} (1,1; {\bf q}, s) \notag\\
G_{yx} (2,2; {\bf q}, s) &=-G_{xy} (1,1; {\bf q}, s).
\end{align}
We thus learn that all nonzero elements of $G_{\alpha \beta}  (\kappa,\kappa';{\bf q}, s)$ are proportional to the electronic valley Chern number, ${\rm sign}(m_0)$. 
When the ${\cal T}$-symmetry is broken by an exchange field, the electronic bands undergo a Zeeman shift. As a consequence, the Dirac mass in both valleys will no longer have the same magnitude \cite{qi2015giant}.
However, as seen from Eq.~(\ref{eq:calG}), the MBC at each valley does not depend on the magnitude of the Dirac mass, but only on its sign. Hence, the total MBC vanishes even in the presence of a valley Zeeman splitting, provided that the later is not strong enough to induce a band inversion in one of the valleys. 

Admittedly, a residual nonzero MBC could  result from the ${\cal T}-$breaking perturbation, as the Dirac velocities as well as the strengths of the electron-phonon coupling in both valleys could be affected differently by $B$, thereby preventing a cancellation between valleys in the overall MBC.
In that case, as per Eq. (\ref{eq:calG}), the resulting molecular Berry curvature would be proportional to the valley Chern number of electrons, this being a generalization of the result found by Saparov et al. in Ref. \cite{saparov2022lattice}.
We have neglected this residual MBC in the treatment of Sec. \ref{sec:TR_breaking}.

\end{widetext}

\bibliography{refs}{}

\begin{thebibliography}{52}%
\makeatletter
\providecommand \@ifxundefined [1]{%
 \@ifx{#1\undefined}
}%
\providecommand \@ifnum [1]{%
 \ifnum #1\expandafter \@firstoftwo
 \else \expandafter \@secondoftwo
 \fi
}%
\providecommand \@ifx [1]{%
 \ifx #1\expandafter \@firstoftwo
 \else \expandafter \@secondoftwo
 \fi
}%
\providecommand \natexlab [1]{#1}%
\providecommand \enquote  [1]{``#1''}%
\providecommand \bibnamefont  [1]{#1}%
\providecommand \bibfnamefont [1]{#1}%
\providecommand \citenamefont [1]{#1}%
\providecommand \href@noop [0]{\@secondoftwo}%
\providecommand \href [0]{\begingroup \@sanitize@url \@href}%
\providecommand \@href[1]{\@@startlink{#1}\@@href}%
\providecommand \@@href[1]{\endgroup#1\@@endlink}%
\providecommand \@sanitize@url [0]{\catcode `\\12\catcode `\$12\catcode
  `\&12\catcode `\#12\catcode `\^12\catcode `\_12\catcode `\%12\relax}%
\providecommand \@@startlink[1]{}%
\providecommand \@@endlink[0]{}%
\providecommand \url  [0]{\begingroup\@sanitize@url \@url }%
\providecommand \@url [1]{\endgroup\@href {#1}{\urlprefix }}%
\providecommand \urlprefix  [0]{URL }%
\providecommand \Eprint [0]{\href }%
\providecommand \doibase [0]{https://doi.org/}%
\providecommand \selectlanguage [0]{\@gobble}%
\providecommand \bibinfo  [0]{\@secondoftwo}%
\providecommand \bibfield  [0]{\@secondoftwo}%
\providecommand \translation [1]{[#1]}%
\providecommand \BibitemOpen [0]{}%
\providecommand \bibitemStop [0]{}%
\providecommand \bibitemNoStop [0]{.\EOS\space}%
\providecommand \EOS [0]{\spacefactor3000\relax}%
\providecommand \BibitemShut  [1]{\csname bibitem#1\endcsname}%
\let\auto@bib@innerbib\@empty
\bibitem [{\citenamefont {Zhang}\ \emph {et~al.}(2010)\citenamefont {Zhang},
  \citenamefont {Ren}, \citenamefont {Wang},\ and\ \citenamefont
  {Li}}]{zhang2010topological}%
  \BibitemOpen
  \bibfield  {author} {\bibinfo {author} {\bibfnamefont {L.}~\bibnamefont
  {Zhang}}, \bibinfo {author} {\bibfnamefont {J.}~\bibnamefont {Ren}}, \bibinfo
  {author} {\bibfnamefont {J.-S.}\ \bibnamefont {Wang}},\ and\ \bibinfo
  {author} {\bibfnamefont {B.}~\bibnamefont {Li}},\ }\bibfield  {title}
  {\bibinfo {title} {{Topological Nature of the Phonon {H}all Effect}},\ }\href
  {https://doi.org/10.1103/PhysRevLett.105.225901} {\bibfield  {journal}
  {\bibinfo  {journal} {Phys. Rev. Lett.}\ }\textbf {\bibinfo {volume} {105}},\
  \bibinfo {pages} {225901} (\bibinfo {year} {2010})}\BibitemShut {NoStop}%
\bibitem [{\citenamefont {Qin}\ \emph {et~al.}(2012)\citenamefont {Qin},
  \citenamefont {Zhou},\ and\ \citenamefont {Shi}}]{qin2012berry}%
  \BibitemOpen
  \bibfield  {author} {\bibinfo {author} {\bibfnamefont {T.}~\bibnamefont
  {Qin}}, \bibinfo {author} {\bibfnamefont {J.}~\bibnamefont {Zhou}},\ and\
  \bibinfo {author} {\bibfnamefont {J.}~\bibnamefont {Shi}},\ }\bibfield
  {title} {\bibinfo {title} {Berry curvature and the phonon {H}all effect},\
  }\href {https://doi.org/10.1103/PhysRevB.86.104305} {\bibfield  {journal}
  {\bibinfo  {journal} {Phys. Rev. B}\ }\textbf {\bibinfo {volume} {86}},\
  \bibinfo {pages} {104305} (\bibinfo {year} {2012})}\BibitemShut {NoStop}%
\bibitem [{\citenamefont {Liu}\ \emph {et~al.}(2017{\natexlab{a}})\citenamefont
  {Liu}, \citenamefont {Xu},\ and\ \citenamefont {Duan}}]{liu2018berry}%
  \BibitemOpen
  \bibfield  {author} {\bibinfo {author} {\bibfnamefont {Y.}~\bibnamefont
  {Liu}}, \bibinfo {author} {\bibfnamefont {Y.}~\bibnamefont {Xu}},\ and\
  \bibinfo {author} {\bibfnamefont {W.}~\bibnamefont {Duan}},\ }\bibfield
  {title} {\bibinfo {title} {Berry phase and topological effects of phonons},\
  }\href {https://doi.org/10.1093/nsr/nwx086} {\bibfield  {journal} {\bibinfo
  {journal} {National Science Review}\ }\textbf {\bibinfo {volume} {5}},\
  \bibinfo {pages} {314} (\bibinfo {year} {2017}{\natexlab{a}})}\BibitemShut
  {NoStop}%
\bibitem [{\citenamefont {Liu}\ \emph {et~al.}(2020)\citenamefont {Liu},
  \citenamefont {Chen},\ and\ \citenamefont {Xu}}]{liu2020topological}%
  \BibitemOpen
  \bibfield  {author} {\bibinfo {author} {\bibfnamefont {Y.}~\bibnamefont
  {Liu}}, \bibinfo {author} {\bibfnamefont {X.}~\bibnamefont {Chen}},\ and\
  \bibinfo {author} {\bibfnamefont {Y.}~\bibnamefont {Xu}},\ }\bibfield
  {title} {\bibinfo {title} {Topological phononics: From fundamental models to
  real materials},\ }\href {https://doi.org/10.1002/adfm.201904784} {\bibfield
  {journal} {\bibinfo  {journal} {Advanced Functional Materials}\ }\textbf
  {\bibinfo {volume} {30}},\ \bibinfo {pages} {1904784} (\bibinfo {year}
  {2020})}\BibitemShut {NoStop}%
\bibitem [{\citenamefont {Ding}\ \emph {et~al.}(2024)\citenamefont {Ding},
  \citenamefont {Zeng}, \citenamefont {Liu}, \citenamefont {Tang},\ and\
  \citenamefont {Chen}}]{ding2024topological}%
  \BibitemOpen
  \bibfield  {author} {\bibinfo {author} {\bibfnamefont {Z.-K.}\ \bibnamefont
  {Ding}}, \bibinfo {author} {\bibfnamefont {Y.-J.}\ \bibnamefont {Zeng}},
  \bibinfo {author} {\bibfnamefont {W.}~\bibnamefont {Liu}}, \bibinfo {author}
  {\bibfnamefont {L.-M.}\ \bibnamefont {Tang}},\ and\ \bibinfo {author}
  {\bibfnamefont {K.-Q.}\ \bibnamefont {Chen}},\ }\bibfield  {title} {\bibinfo
  {title} {Topological phonons and thermoelectric conversion in crystalline
  materials},\ }\href {https://doi.org/https://doi.org/10.1002/adfm.202401684}
  {\bibfield  {journal} {\bibinfo  {journal} {Advanced Functional Materials}\
  }\textbf {\bibinfo {volume} {34}},\ \bibinfo {pages} {2401684} (\bibinfo
  {year} {2024})}\BibitemShut {NoStop}%
\bibitem [{\citenamefont {Zhang}\ \emph {et~al.}(2018)\citenamefont {Zhang},
  \citenamefont {Song}, \citenamefont {Alexandradinata}, \citenamefont {Weng},
  \citenamefont {Fang}, \citenamefont {Lu},\ and\ \citenamefont
  {Fang}}]{zhang2018double}%
  \BibitemOpen
  \bibfield  {author} {\bibinfo {author} {\bibfnamefont {T.}~\bibnamefont
  {Zhang}}, \bibinfo {author} {\bibfnamefont {Z.}~\bibnamefont {Song}},
  \bibinfo {author} {\bibfnamefont {A.}~\bibnamefont {Alexandradinata}},
  \bibinfo {author} {\bibfnamefont {H.}~\bibnamefont {Weng}}, \bibinfo {author}
  {\bibfnamefont {C.}~\bibnamefont {Fang}}, \bibinfo {author} {\bibfnamefont
  {L.}~\bibnamefont {Lu}},\ and\ \bibinfo {author} {\bibfnamefont
  {Z.}~\bibnamefont {Fang}},\ }\bibfield  {title} {\bibinfo {title}
  {Double-{W}eyl phonons in transition-metal monosilicides},\ }\href
  {https://doi.org/10.1103/PhysRevLett.120.016401} {\bibfield  {journal}
  {\bibinfo  {journal} {Phys. Rev. Lett.}\ }\textbf {\bibinfo {volume} {120}},\
  \bibinfo {pages} {016401} (\bibinfo {year} {2018})}\BibitemShut {NoStop}%
\bibitem [{\citenamefont {Miao}\ \emph {et~al.}(2018)\citenamefont {Miao},
  \citenamefont {Zhang}, \citenamefont {Wang}, \citenamefont {Meyers},
  \citenamefont {Said}, \citenamefont {Wang}, \citenamefont {Shi},
  \citenamefont {Weng}, \citenamefont {Fang},\ and\ \citenamefont
  {Dean}}]{miao1018observation}%
  \BibitemOpen
  \bibfield  {author} {\bibinfo {author} {\bibfnamefont {H.}~\bibnamefont
  {Miao}}, \bibinfo {author} {\bibfnamefont {T.~T.}\ \bibnamefont {Zhang}},
  \bibinfo {author} {\bibfnamefont {L.}~\bibnamefont {Wang}}, \bibinfo {author}
  {\bibfnamefont {D.}~\bibnamefont {Meyers}}, \bibinfo {author} {\bibfnamefont
  {A.~H.}\ \bibnamefont {Said}}, \bibinfo {author} {\bibfnamefont {Y.~L.}\
  \bibnamefont {Wang}}, \bibinfo {author} {\bibfnamefont {Y.~G.}\ \bibnamefont
  {Shi}}, \bibinfo {author} {\bibfnamefont {H.~M.}\ \bibnamefont {Weng}},
  \bibinfo {author} {\bibfnamefont {Z.}~\bibnamefont {Fang}},\ and\ \bibinfo
  {author} {\bibfnamefont {M.~P.~M.}\ \bibnamefont {Dean}},\ }\bibfield
  {title} {\bibinfo {title} {Observation of double {W}eyl phonons in
  parity-breaking {FeSi}},\ }\href
  {https://doi.org/10.1103/PhysRevLett.121.035302} {\bibfield  {journal}
  {\bibinfo  {journal} {Phys. Rev. Lett.}\ }\textbf {\bibinfo {volume} {121}},\
  \bibinfo {pages} {035302} (\bibinfo {year} {2018})}\BibitemShut {NoStop}%
\bibitem [{\citenamefont {Jin}\ \emph {et~al.}(2018)\citenamefont {Jin},
  \citenamefont {Wang},\ and\ \citenamefont {Xu}}]{jin2018recipe}%
  \BibitemOpen
  \bibfield  {author} {\bibinfo {author} {\bibfnamefont {Y.}~\bibnamefont
  {Jin}}, \bibinfo {author} {\bibfnamefont {R.}~\bibnamefont {Wang}},\ and\
  \bibinfo {author} {\bibfnamefont {H.}~\bibnamefont {Xu}},\ }\bibfield
  {title} {\bibinfo {title} {{Recipe for Dirac phonon states with a quantized
  valley Berry phase in two-dimensional hexagonal lattices}},\ }\href
  {https://doi.org/10.1021/acs.nanolett.8b03492} {\bibfield  {journal}
  {\bibinfo  {journal} {Nano letters}\ }\textbf {\bibinfo {volume} {18}},\
  \bibinfo {pages} {7755} (\bibinfo {year} {2018})}\BibitemShut {NoStop}%
\bibitem [{\citenamefont {Coh}(2023)}]{coh2023classification}%
  \BibitemOpen
  \bibfield  {author} {\bibinfo {author} {\bibfnamefont {S.}~\bibnamefont
  {Coh}},\ }\bibfield  {title} {\bibinfo {title} {Classification of materials
  with phonon angular momentum and microscopic origin of angular momentum},\
  }\href {https://doi.org/10.1103/PhysRevB.108.134307} {\bibfield  {journal}
  {\bibinfo  {journal} {Phys. Rev. B}\ }\textbf {\bibinfo {volume} {108}},\
  \bibinfo {pages} {134307} (\bibinfo {year} {2023})}\BibitemShut {NoStop}%
\bibitem [{\citenamefont {Xu}\ \emph {et~al.}(2024)\citenamefont {Xu},
  \citenamefont {Vergniory}, \citenamefont {Ma}, \citenamefont {Ma{\~n}es},
  \citenamefont {Song}, \citenamefont {Bernevig}, \citenamefont {Regnault},\
  and\ \citenamefont {Elcoro}}]{xu2024catalog}%
  \BibitemOpen
  \bibfield  {author} {\bibinfo {author} {\bibfnamefont {Y.}~\bibnamefont
  {Xu}}, \bibinfo {author} {\bibfnamefont {M.}~\bibnamefont {Vergniory}},
  \bibinfo {author} {\bibfnamefont {D.-S.}\ \bibnamefont {Ma}}, \bibinfo
  {author} {\bibfnamefont {J.~L.}\ \bibnamefont {Ma{\~n}es}}, \bibinfo {author}
  {\bibfnamefont {Z.-D.}\ \bibnamefont {Song}}, \bibinfo {author}
  {\bibfnamefont {B.~A.}\ \bibnamefont {Bernevig}}, \bibinfo {author}
  {\bibfnamefont {N.}~\bibnamefont {Regnault}},\ and\ \bibinfo {author}
  {\bibfnamefont {L.}~\bibnamefont {Elcoro}},\ }\bibfield  {title} {\bibinfo
  {title} {Catalog of topological phonon materials},\ }\href
  {https://doi.org/10.1126/science.adf8458} {\bibfield  {journal} {\bibinfo
  {journal} {Science}\ }\textbf {\bibinfo {volume} {384}},\ \bibinfo {pages}
  {eadf8458} (\bibinfo {year} {2024})}\BibitemShut {NoStop}%
\bibitem [{\citenamefont {Xiao}\ \emph {et~al.}(2010)\citenamefont {Xiao},
  \citenamefont {Chang},\ and\ \citenamefont {Niu}}]{xiao2010berry}%
  \BibitemOpen
  \bibfield  {author} {\bibinfo {author} {\bibfnamefont {D.}~\bibnamefont
  {Xiao}}, \bibinfo {author} {\bibfnamefont {M.-C.}\ \bibnamefont {Chang}},\
  and\ \bibinfo {author} {\bibfnamefont {Q.}~\bibnamefont {Niu}},\ }\bibfield
  {title} {\bibinfo {title} {{Berry phase effects on electronic properties}},\
  }\href {https://doi.org/10.1103/RevModPhys.82.1959} {\bibfield  {journal}
  {\bibinfo  {journal} {Rev. Mod. Phys.}\ }\textbf {\bibinfo {volume} {82}},\
  \bibinfo {pages} {1959} (\bibinfo {year} {2010})}\BibitemShut {NoStop}%
\bibitem [{\citenamefont {Liu}\ \emph {et~al.}(2017{\natexlab{b}})\citenamefont
  {Liu}, \citenamefont {Xu}, \citenamefont {Zhang},\ and\ \citenamefont
  {Duan}}]{liu2017model}%
  \BibitemOpen
  \bibfield  {author} {\bibinfo {author} {\bibfnamefont {Y.}~\bibnamefont
  {Liu}}, \bibinfo {author} {\bibfnamefont {Y.}~\bibnamefont {Xu}}, \bibinfo
  {author} {\bibfnamefont {S.-C.}\ \bibnamefont {Zhang}},\ and\ \bibinfo
  {author} {\bibfnamefont {W.}~\bibnamefont {Duan}},\ }\bibfield  {title}
  {\bibinfo {title} {Model for topological phononics and phonon diode},\ }\href
  {https://doi.org/10.1103/PhysRevB.96.064106} {\bibfield  {journal} {\bibinfo
  {journal} {Phys. Rev. B}\ }\textbf {\bibinfo {volume} {96}},\ \bibinfo
  {pages} {064106} (\bibinfo {year} {2017}{\natexlab{b}})}\BibitemShut
  {NoStop}%
\bibitem [{\citenamefont {Saito}\ \emph {et~al.}(2019)\citenamefont {Saito},
  \citenamefont {Misaki}, \citenamefont {Ishizuka},\ and\ \citenamefont
  {Nagaosa}}]{saito2019berry}%
  \BibitemOpen
  \bibfield  {author} {\bibinfo {author} {\bibfnamefont {T.}~\bibnamefont
  {Saito}}, \bibinfo {author} {\bibfnamefont {K.}~\bibnamefont {Misaki}},
  \bibinfo {author} {\bibfnamefont {H.}~\bibnamefont {Ishizuka}},\ and\
  \bibinfo {author} {\bibfnamefont {N.}~\bibnamefont {Nagaosa}},\ }\bibfield
  {title} {\bibinfo {title} {Berry phase of phonons and thermal {H}all effect
  in nonmagnetic insulators},\ }\href
  {https://doi.org/10.1103/PhysRevLett.123.255901} {\bibfield  {journal}
  {\bibinfo  {journal} {Phys. Rev. Lett.}\ }\textbf {\bibinfo {volume} {123}},\
  \bibinfo {pages} {255901} (\bibinfo {year} {2019})}\BibitemShut {NoStop}%
\bibitem [{\citenamefont {Saparov}\ \emph {et~al.}(2022)\citenamefont
  {Saparov}, \citenamefont {Xiong}, \citenamefont {Ren},\ and\ \citenamefont
  {Niu}}]{saparov2022lattice}%
  \BibitemOpen
  \bibfield  {author} {\bibinfo {author} {\bibfnamefont {D.}~\bibnamefont
  {Saparov}}, \bibinfo {author} {\bibfnamefont {B.}~\bibnamefont {Xiong}},
  \bibinfo {author} {\bibfnamefont {Y.}~\bibnamefont {Ren}},\ and\ \bibinfo
  {author} {\bibfnamefont {Q.}~\bibnamefont {Niu}},\ }\bibfield  {title}
  {\bibinfo {title} {{Lattice dynamics with molecular Berry curvature: Chiral
  optical phonons}},\ }\href {https://doi.org/10.1103/PhysRevB.105.064303}
  {\bibfield  {journal} {\bibinfo  {journal} {Phys. Rev. B}\ }\textbf {\bibinfo
  {volume} {105}},\ \bibinfo {pages} {064303} (\bibinfo {year}
  {2022})}\BibitemShut {NoStop}%
\bibitem [{\citenamefont {Liu}\ \emph {et~al.}(2021)\citenamefont {Liu},
  \citenamefont {Yu}, \citenamefont {Ning}, \citenamefont {Yi}, \citenamefont
  {Miao}, \citenamefont {Min}, \citenamefont {Zhao}, \citenamefont {Ning},
  \citenamefont {Lopez}, \citenamefont {Zhu} \emph {et~al.}}]{liu2021spin}%
  \BibitemOpen
  \bibfield  {author} {\bibinfo {author} {\bibfnamefont {J.}~\bibnamefont
  {Liu}}, \bibinfo {author} {\bibfnamefont {J.}~\bibnamefont {Yu}}, \bibinfo
  {author} {\bibfnamefont {J.~L.}\ \bibnamefont {Ning}}, \bibinfo {author}
  {\bibfnamefont {H.}~\bibnamefont {Yi}}, \bibinfo {author} {\bibfnamefont
  {L.}~\bibnamefont {Miao}}, \bibinfo {author} {\bibfnamefont {L.}~\bibnamefont
  {Min}}, \bibinfo {author} {\bibfnamefont {Y.}~\bibnamefont {Zhao}}, \bibinfo
  {author} {\bibfnamefont {W.}~\bibnamefont {Ning}}, \bibinfo {author}
  {\bibfnamefont {K.}~\bibnamefont {Lopez}}, \bibinfo {author} {\bibfnamefont
  {Y.}~\bibnamefont {Zhu}}, \emph {et~al.},\ }\bibfield  {title} {\bibinfo
  {title} {{Spin-valley locking and bulk quantum {H}all effect in a
  noncentrosymmetric Dirac semimetal {BaMnSb$_2$}}},\ }\href
  {https://doi.org/https://doi.org/10.1038/s41467-021-24369-1} {\bibfield
  {journal} {\bibinfo  {journal} {Nature communications}\ }\textbf {\bibinfo
  {volume} {12}},\ \bibinfo {pages} {4062} (\bibinfo {year}
  {2021})}\BibitemShut {NoStop}%
\bibitem [{\citenamefont {Hu}\ \emph {et~al.}(2021)\citenamefont {Hu},
  \citenamefont {Yu}, \citenamefont {Garate},\ and\ \citenamefont
  {Liu}}]{hu2021phonon}%
  \BibitemOpen
  \bibfield  {author} {\bibinfo {author} {\bibfnamefont {L.-H.}\ \bibnamefont
  {Hu}}, \bibinfo {author} {\bibfnamefont {J.}~\bibnamefont {Yu}}, \bibinfo
  {author} {\bibfnamefont {I.}~\bibnamefont {Garate}},\ and\ \bibinfo {author}
  {\bibfnamefont {C.-X.}\ \bibnamefont {Liu}},\ }\bibfield  {title} {\bibinfo
  {title} {{Phonon Helicity Induced by Electronic Berry Curvature in Dirac
  Materials}},\ }\href {https://doi.org/10.1103/PhysRevLett.127.125901}
  {\bibfield  {journal} {\bibinfo  {journal} {Phys. Rev. Lett.}\ }\textbf
  {\bibinfo {volume} {127}},\ \bibinfo {pages} {125901} (\bibinfo {year}
  {2021})}\BibitemShut {NoStop}%
\bibitem [{Note1()}]{Note1}%
  \BibitemOpen
  \bibinfo {note} {Along a loosely related line of inquiry, Ref. \cite
  {lin2024quantum} has shown that the quantum oscillations in acoustic phonons
  are altered by the electronic Berry phase.}\BibitemShut {Stop}%
\bibitem [{\citenamefont {Hamada}\ \emph {et~al.}(2018)\citenamefont {Hamada},
  \citenamefont {Minamitani}, \citenamefont {Hirayama},\ and\ \citenamefont
  {Murakami}}]{hamada2018phonon}%
  \BibitemOpen
  \bibfield  {author} {\bibinfo {author} {\bibfnamefont {M.}~\bibnamefont
  {Hamada}}, \bibinfo {author} {\bibfnamefont {E.}~\bibnamefont {Minamitani}},
  \bibinfo {author} {\bibfnamefont {M.}~\bibnamefont {Hirayama}},\ and\
  \bibinfo {author} {\bibfnamefont {S.}~\bibnamefont {Murakami}},\ }\bibfield
  {title} {\bibinfo {title} {Phonon angular momentum induced by the temperature
  gradient},\ }\href {https://doi.org/10.1103/PhysRevLett.121.175301}
  {\bibfield  {journal} {\bibinfo  {journal} {Phys. Rev. Lett.}\ }\textbf
  {\bibinfo {volume} {121}},\ \bibinfo {pages} {175301} (\bibinfo {year}
  {2018})}\BibitemShut {NoStop}%
\bibitem [{\citenamefont {Zhang}\ \emph {et~al.}(2024)\citenamefont {Zhang},
  \citenamefont {Peshcherenko}, \citenamefont {Yang}, \citenamefont {Ward},
  \citenamefont {Raghuvanshi}, \citenamefont {Lindsay}, \citenamefont {Felser},
  \citenamefont {Zhang}, \citenamefont {Yan},\ and\ \citenamefont
  {Miao}}]{zhang2024observation}%
  \BibitemOpen
  \bibfield  {author} {\bibinfo {author} {\bibfnamefont {H.}~\bibnamefont
  {Zhang}}, \bibinfo {author} {\bibfnamefont {N.}~\bibnamefont {Peshcherenko}},
  \bibinfo {author} {\bibfnamefont {F.}~\bibnamefont {Yang}}, \bibinfo {author}
  {\bibfnamefont {T.}~\bibnamefont {Ward}}, \bibinfo {author} {\bibfnamefont
  {P.}~\bibnamefont {Raghuvanshi}}, \bibinfo {author} {\bibfnamefont
  {L.}~\bibnamefont {Lindsay}}, \bibinfo {author} {\bibfnamefont
  {C.}~\bibnamefont {Felser}}, \bibinfo {author} {\bibfnamefont
  {Y.}~\bibnamefont {Zhang}}, \bibinfo {author} {\bibfnamefont {J.-Q.}\
  \bibnamefont {Yan}},\ and\ \bibinfo {author} {\bibfnamefont {H.}~\bibnamefont
  {Miao}},\ }\bibfield  {title} {\bibinfo {title} {Observation of phonon
  angular momentum},\ }\href {https://doi.org/10.48550/arXiv.2409.13462}
  {\bibfield  {journal} {\bibinfo  {journal} {arXiv preprint arXiv:2409.13462}\
  } (\bibinfo {year} {2024})}\BibitemShut {NoStop}%
\bibitem [{\citenamefont {Parlak}\ \emph {et~al.}(2023)\citenamefont {Parlak},
  \citenamefont {Ghosh},\ and\ \citenamefont {Garate}}]{parlak2023detection}%
  \BibitemOpen
  \bibfield  {author} {\bibinfo {author} {\bibfnamefont {S.}~\bibnamefont
  {Parlak}}, \bibinfo {author} {\bibfnamefont {S.}~\bibnamefont {Ghosh}},\ and\
  \bibinfo {author} {\bibfnamefont {I.}~\bibnamefont {Garate}},\ }\bibfield
  {title} {\bibinfo {title} {{Detection of phonon helicity in nonchiral
  crystals with Raman scattering}},\ }\href
  {https://doi.org/10.1103/PhysRevB.107.104308} {\bibfield  {journal} {\bibinfo
   {journal} {Phys. Rev. B}\ }\textbf {\bibinfo {volume} {107}},\ \bibinfo
  {pages} {104308} (\bibinfo {year} {2023})}\BibitemShut {NoStop}%
\bibitem [{\citenamefont {Chen}\ \emph {et~al.}(2024)\citenamefont {Chen},
  \citenamefont {Jin}, \citenamefont {Liao},\ and\ \citenamefont
  {Mu}}]{chen2024thermoelectric}%
  \BibitemOpen
  \bibfield  {author} {\bibinfo {author} {\bibfnamefont {Y.}~\bibnamefont
  {Chen}}, \bibinfo {author} {\bibfnamefont {R.}~\bibnamefont {Jin}}, \bibinfo
  {author} {\bibfnamefont {B.}~\bibnamefont {Liao}},\ and\ \bibinfo {author}
  {\bibfnamefont {S.}~\bibnamefont {Mu}},\ }\bibfield  {title} {\bibinfo
  {title} {Thermoelectric transport in {W}eyl semimetal
  {$\mathrm{Ba}\mathrm{Mn}{\mathrm{Sb}}_{2}$}: A first-principles study},\
  }\href {https://doi.org/10.1103/PhysRevMaterials.8.085401} {\bibfield
  {journal} {\bibinfo  {journal} {Phys. Rev. Mater.}\ }\textbf {\bibinfo
  {volume} {8}},\ \bibinfo {pages} {085401} (\bibinfo {year}
  {2024})}\BibitemShut {NoStop}%
\bibitem [{\citenamefont {Michel}\ and\ \citenamefont
  {Verberck}(2011)}]{michel2011phonon}%
  \BibitemOpen
  \bibfield  {author} {\bibinfo {author} {\bibfnamefont {K.~H.}\ \bibnamefont
  {Michel}}\ and\ \bibinfo {author} {\bibfnamefont {B.}~\bibnamefont
  {Verberck}},\ }\bibfield  {title} {\bibinfo {title} {{Phonon dispersions and
  piezoelectricity in bulk and multilayers of hexagonal boron nitride}},\
  }\href {https://doi.org/10.1103/PhysRevB.83.115328} {\bibfield  {journal}
  {\bibinfo  {journal} {Phys. Rev. B}\ }\textbf {\bibinfo {volume} {83}},\
  \bibinfo {pages} {115328} (\bibinfo {year} {2011})}\BibitemShut {NoStop}%
\bibitem [{\citenamefont {Maradudin}\ and\ \citenamefont
  {Vosko}(1968)}]{maradudin1968symmetry}%
  \BibitemOpen
  \bibfield  {author} {\bibinfo {author} {\bibfnamefont {A.~A.}\ \bibnamefont
  {Maradudin}}\ and\ \bibinfo {author} {\bibfnamefont {S.~H.}\ \bibnamefont
  {Vosko}},\ }\bibfield  {title} {\bibinfo {title} {Symmetry properties of the
  normal vibrations of a crystal},\ }\href
  {https://doi.org/10.1103/RevModPhys.40.1} {\bibfield  {journal} {\bibinfo
  {journal} {Rev. Mod. Phys.}\ }\textbf {\bibinfo {volume} {40}},\ \bibinfo
  {pages} {1} (\bibinfo {year} {1968})}\BibitemShut {NoStop}%
\bibitem [{\citenamefont {Sun}\ \emph {et~al.}(2020)\citenamefont {Sun},
  \citenamefont {Gao},\ and\ \citenamefont {Wang}}]{sun2012current}%
  \BibitemOpen
  \bibfield  {author} {\bibinfo {author} {\bibfnamefont {K.}~\bibnamefont
  {Sun}}, \bibinfo {author} {\bibfnamefont {Z.}~\bibnamefont {Gao}},\ and\
  \bibinfo {author} {\bibfnamefont {J.-S.}\ \bibnamefont {Wang}},\ }\bibfield
  {title} {\bibinfo {title} {{Current-induced phonon {H}all effect}},\ }\href
  {https://doi.org/10.1103/PhysRevB.102.134311} {\bibfield  {journal} {\bibinfo
   {journal} {Phys. Rev. B}\ }\textbf {\bibinfo {volume} {102}},\ \bibinfo
  {pages} {134311} (\bibinfo {year} {2020})}\BibitemShut {NoStop}%
\bibitem [{Note2()}]{Note2}%
  \BibitemOpen
  \bibinfo {note} {We make the gauge choice to place the phase factor $\protect
  \qopname \relax o{exp}(\pm i\phi )$ always in front of $\protect \qopname
  \relax o{sin}(\theta /2)$. We could equally well have chosen to place the
  phase factors always in front of $\protect \qopname \relax o{cos}(\theta
  /2)$. Either choice will ensure that $|1\rangle \langle 2|$ has no
  discontinuities as the wave vector is varied ($|1\rangle \langle 1|$ and
  $|2\rangle \langle 2|$ already have that property irrespective of the gauge
  choice).}\BibitemShut {Stop}%
\bibitem [{\citenamefont {Sticlet}\ \emph {et~al.}(2012)\citenamefont
  {Sticlet}, \citenamefont {Pi\'echon}, \citenamefont {Fuchs}, \citenamefont
  {Kalugin},\ and\ \citenamefont {Simon}}]{sticlet2012geometrical}%
  \BibitemOpen
  \bibfield  {author} {\bibinfo {author} {\bibfnamefont {D.}~\bibnamefont
  {Sticlet}}, \bibinfo {author} {\bibfnamefont {F.}~\bibnamefont {Pi\'echon}},
  \bibinfo {author} {\bibfnamefont {J.-N.}\ \bibnamefont {Fuchs}}, \bibinfo
  {author} {\bibfnamefont {P.}~\bibnamefont {Kalugin}},\ and\ \bibinfo {author}
  {\bibfnamefont {P.}~\bibnamefont {Simon}},\ }\bibfield  {title} {\bibinfo
  {title} {{Geometrical engineering of a two-band Chern insulator in two
  dimensions with arbitrary topological index}},\ }\href
  {https://doi.org/10.1103/PhysRevB.85.165456} {\bibfield  {journal} {\bibinfo
  {journal} {Phys. Rev. B}\ }\textbf {\bibinfo {volume} {85}},\ \bibinfo
  {pages} {165456} (\bibinfo {year} {2012})}\BibitemShut {NoStop}%
\bibitem [{Note3()}]{Note3}%
  \BibitemOpen
  \bibinfo {note} {In the presence of time-reversal symmetry, the eigenvectors
  of ${\protect \cal H}_{\protect \rm ph}$ coincide with those of ${\protect
  \cal D}$. Consequently, the Berry curvature can be calculated directly from
  the dynamical matrix without having to deal with the Hamiltonian. However, in
  the end of the paper we will consider the case of broken time-reversal
  symmetry. In that case, the effective phonon Hamiltonian is no longer given
  by the square root of the dynamical matrix, and its eigenvectors no longer
  coincide with those of the dynamical matrix. In that more general case, we
  can still obtain the phonon Berry curvature from Eq.~(\ref
  {eq:BC}).}\BibitemShut {Stop}%
\bibitem [{\citenamefont {Rinkel}\ \emph {et~al.}(2017)\citenamefont {Rinkel},
  \citenamefont {Lopes},\ and\ \citenamefont {Garate}}]{rinkel2017signatures}%
  \BibitemOpen
  \bibfield  {author} {\bibinfo {author} {\bibfnamefont {P.}~\bibnamefont
  {Rinkel}}, \bibinfo {author} {\bibfnamefont {P.~L.~S.}\ \bibnamefont
  {Lopes}},\ and\ \bibinfo {author} {\bibfnamefont {I.}~\bibnamefont
  {Garate}},\ }\bibfield  {title} {\bibinfo {title} {{Signatures of the Chiral
  Anomaly in Phonon Dynamics}},\ }\href
  {https://doi.org/10.1103/PhysRevLett.119.107401} {\bibfield  {journal}
  {\bibinfo  {journal} {Phys. Rev. Lett.}\ }\textbf {\bibinfo {volume} {119}},\
  \bibinfo {pages} {107401} (\bibinfo {year} {2017})}\BibitemShut {NoStop}%
\bibitem [{\citenamefont {Saha}\ \emph {et~al.}(2015)\citenamefont {Saha},
  \citenamefont {L\'egar\'e},\ and\ \citenamefont
  {Garate}}]{saha2015detecting}%
  \BibitemOpen
  \bibfield  {author} {\bibinfo {author} {\bibfnamefont {K.}~\bibnamefont
  {Saha}}, \bibinfo {author} {\bibfnamefont {K.}~\bibnamefont {L\'egar\'e}},\
  and\ \bibinfo {author} {\bibfnamefont {I.}~\bibnamefont {Garate}},\
  }\bibfield  {title} {\bibinfo {title} {Detecting band inversions by measuring
  the environment: Fingerprints of electronic band topology in bulk phonon
  linewidths},\ }\href {https://doi.org/10.1103/PhysRevLett.115.176405}
  {\bibfield  {journal} {\bibinfo  {journal} {Phys. Rev. Lett.}\ }\textbf
  {\bibinfo {volume} {115}},\ \bibinfo {pages} {176405} (\bibinfo {year}
  {2015})}\BibitemShut {NoStop}%
\bibitem [{\citenamefont {Mahan}(2013)}]{mahan2013many}%
  \BibitemOpen
  \bibfield  {author} {\bibinfo {author} {\bibfnamefont {G.~D.}\ \bibnamefont
  {Mahan}},\ }\href@noop {} {\emph {\bibinfo {title} {Many-particle physics}}}\
  (\bibinfo  {publisher} {Springer Science \& Business Media},\ \bibinfo {year}
  {2013})\BibitemShut {NoStop}%
\bibitem [{\citenamefont {Vogl}(1976)}]{vogl1976microscopic}%
  \BibitemOpen
  \bibfield  {author} {\bibinfo {author} {\bibfnamefont {P.}~\bibnamefont
  {Vogl}},\ }\bibfield  {title} {\bibinfo {title} {Microscopic theory of
  electron-phonon interaction in insulators or semiconductors},\ }\href
  {https://doi.org/10.1103/PhysRevB.13.694} {\bibfield  {journal} {\bibinfo
  {journal} {Phys. Rev. B}\ }\textbf {\bibinfo {volume} {13}},\ \bibinfo
  {pages} {694} (\bibinfo {year} {1976})}\BibitemShut {NoStop}%
\bibitem [{\citenamefont {Kaasbjerg}\ \emph {et~al.}(2012)\citenamefont
  {Kaasbjerg}, \citenamefont {Thygesen},\ and\ \citenamefont
  {Jacobsen}}]{kaasbjerg2012phonon}%
  \BibitemOpen
  \bibfield  {author} {\bibinfo {author} {\bibfnamefont {K.}~\bibnamefont
  {Kaasbjerg}}, \bibinfo {author} {\bibfnamefont {K.~S.}\ \bibnamefont
  {Thygesen}},\ and\ \bibinfo {author} {\bibfnamefont {K.~W.}\ \bibnamefont
  {Jacobsen}},\ }\bibfield  {title} {\bibinfo {title} {{Phonon-limited mobility
  in $n$-type single-layer MoS${}_{2}$ from first principles}},\ }\href
  {https://doi.org/10.1103/PhysRevB.85.115317} {\bibfield  {journal} {\bibinfo
  {journal} {Phys. Rev. B}\ }\textbf {\bibinfo {volume} {85}},\ \bibinfo
  {pages} {115317} (\bibinfo {year} {2012})}\BibitemShut {NoStop}%
\bibitem [{\citenamefont {Song}\ and\ \citenamefont
  {Dery}(2013)}]{song2013transport}%
  \BibitemOpen
  \bibfield  {author} {\bibinfo {author} {\bibfnamefont {Y.}~\bibnamefont
  {Song}}\ and\ \bibinfo {author} {\bibfnamefont {H.}~\bibnamefont {Dery}},\
  }\bibfield  {title} {\bibinfo {title} {Transport theory of monolayer
  transition-metal dichalcogenides through symmetry},\ }\href
  {https://doi.org/10.1103/PhysRevLett.111.026601} {\bibfield  {journal}
  {\bibinfo  {journal} {Phys. Rev. Lett.}\ }\textbf {\bibinfo {volume} {111}},\
  \bibinfo {pages} {026601} (\bibinfo {year} {2013})}\BibitemShut {NoStop}%
\bibitem [{\citenamefont {Giustino}(2017)}]{giustino2017electron}%
  \BibitemOpen
  \bibfield  {author} {\bibinfo {author} {\bibfnamefont {F.}~\bibnamefont
  {Giustino}},\ }\bibfield  {title} {\bibinfo {title} {Electron-phonon
  interactions from first principles},\ }\href
  {https://doi.org/10.1103/RevModPhys.89.015003} {\bibfield  {journal}
  {\bibinfo  {journal} {Rev. Mod. Phys.}\ }\textbf {\bibinfo {volume} {89}},\
  \bibinfo {pages} {015003} (\bibinfo {year} {2017})}\BibitemShut {NoStop}%
\bibitem [{\citenamefont {Sakai}\ \emph {et~al.}(2020)\citenamefont {Sakai},
  \citenamefont {Fujimura}, \citenamefont {Sakuragi}, \citenamefont {Ochi},
  \citenamefont {Kurihara}, \citenamefont {Miyake}, \citenamefont {Tokunaga},
  \citenamefont {Kojima}, \citenamefont {Hashizume}, \citenamefont {Muro},
  \citenamefont {Kuroda}, \citenamefont {Kondo}, \citenamefont {Kida},
  \citenamefont {Hagiwara}, \citenamefont {Kuroki}, \citenamefont {Kondo},
  \citenamefont {Tsuruda}, \citenamefont {Murakawa},\ and\ \citenamefont
  {Hanasaki}}]{sakai2020bulk}%
  \BibitemOpen
  \bibfield  {author} {\bibinfo {author} {\bibfnamefont {H.}~\bibnamefont
  {Sakai}}, \bibinfo {author} {\bibfnamefont {H.}~\bibnamefont {Fujimura}},
  \bibinfo {author} {\bibfnamefont {S.}~\bibnamefont {Sakuragi}}, \bibinfo
  {author} {\bibfnamefont {M.}~\bibnamefont {Ochi}}, \bibinfo {author}
  {\bibfnamefont {R.}~\bibnamefont {Kurihara}}, \bibinfo {author}
  {\bibfnamefont {A.}~\bibnamefont {Miyake}}, \bibinfo {author} {\bibfnamefont
  {M.}~\bibnamefont {Tokunaga}}, \bibinfo {author} {\bibfnamefont
  {T.}~\bibnamefont {Kojima}}, \bibinfo {author} {\bibfnamefont
  {D.}~\bibnamefont {Hashizume}}, \bibinfo {author} {\bibfnamefont
  {T.}~\bibnamefont {Muro}}, \bibinfo {author} {\bibfnamefont {K.}~\bibnamefont
  {Kuroda}}, \bibinfo {author} {\bibfnamefont {T.}~\bibnamefont {Kondo}},
  \bibinfo {author} {\bibfnamefont {T.}~\bibnamefont {Kida}}, \bibinfo {author}
  {\bibfnamefont {M.}~\bibnamefont {Hagiwara}}, \bibinfo {author}
  {\bibfnamefont {K.}~\bibnamefont {Kuroki}}, \bibinfo {author} {\bibfnamefont
  {M.}~\bibnamefont {Kondo}}, \bibinfo {author} {\bibfnamefont
  {K.}~\bibnamefont {Tsuruda}}, \bibinfo {author} {\bibfnamefont
  {H.}~\bibnamefont {Murakawa}},\ and\ \bibinfo {author} {\bibfnamefont
  {N.}~\bibnamefont {Hanasaki}},\ }\bibfield  {title} {\bibinfo {title} {Bulk
  quantum hall effect of spin-valley coupled dirac fermions in the polar
  antiferromagnet {${\mathrm{BaMnSb}}_{2}$}},\ }\href
  {https://doi.org/10.1103/PhysRevB.101.081104} {\bibfield  {journal} {\bibinfo
   {journal} {Phys. Rev. B}\ }\textbf {\bibinfo {volume} {101}},\ \bibinfo
  {pages} {081104} (\bibinfo {year} {2020})}\BibitemShut {NoStop}%
\bibitem [{Note4()}]{Note4}%
  \BibitemOpen
  \bibinfo {note} {For example, the nonlinear thermal Hall effect might be
  sensitive to the phonon Berry curvature in ${\protect \cal T}$-preserving
  systems; see e.g. Ref. \cite {varshney2023intrinsic}. We will not explore
  such possibility in the present work since the effect may be hardly
  measurable, as the linear phonon thermal Hall conductivity that is being
  measured nowadays is already small.}\BibitemShut {Stop}%
\bibitem [{Note5()}]{Note5}%
  \BibitemOpen
  \bibinfo {note} {The spin susceptibility in this regime originates from
  interband matrix elements of the spin operator and is proportional to the
  square of the wave vector of the applied perturbation. It therefore vanishes
  for a spatially uniform $B$.}\BibitemShut {Stop}%
\bibitem [{\citenamefont {Qi}\ \emph {et~al.}(2015)\citenamefont {Qi},
  \citenamefont {Li}, \citenamefont {Niu},\ and\ \citenamefont
  {Feng}}]{qi2015giant}%
  \BibitemOpen
  \bibfield  {author} {\bibinfo {author} {\bibfnamefont {J.}~\bibnamefont
  {Qi}}, \bibinfo {author} {\bibfnamefont {X.}~\bibnamefont {Li}}, \bibinfo
  {author} {\bibfnamefont {Q.}~\bibnamefont {Niu}},\ and\ \bibinfo {author}
  {\bibfnamefont {J.}~\bibnamefont {Feng}},\ }\bibfield  {title} {\bibinfo
  {title} {{Giant and tunable valley degeneracy splitting in
  ${\mathrm{MoTe}}_{2}$}},\ }\href {https://doi.org/10.1103/PhysRevB.92.121403}
  {\bibfield  {journal} {\bibinfo  {journal} {Phys. Rev. B}\ }\textbf {\bibinfo
  {volume} {92}},\ \bibinfo {pages} {121403} (\bibinfo {year}
  {2015})}\BibitemShut {NoStop}%
\bibitem [{\citenamefont {Sun}\ \emph {et~al.}(2021)\citenamefont {Sun},
  \citenamefont {Gao},\ and\ \citenamefont {Wang}}]{sun2021phonon}%
  \BibitemOpen
  \bibfield  {author} {\bibinfo {author} {\bibfnamefont {K.}~\bibnamefont
  {Sun}}, \bibinfo {author} {\bibfnamefont {Z.}~\bibnamefont {Gao}},\ and\
  \bibinfo {author} {\bibfnamefont {J.-S.}\ \bibnamefont {Wang}},\ }\bibfield
  {title} {\bibinfo {title} {Phonon {H}all effect with first-principles
  calculations},\ }\href {https://doi.org/10.1103/PhysRevB.103.214301}
  {\bibfield  {journal} {\bibinfo  {journal} {Phys. Rev. B}\ }\textbf {\bibinfo
  {volume} {103}},\ \bibinfo {pages} {214301} (\bibinfo {year}
  {2021})}\BibitemShut {NoStop}%
\bibitem [{\citenamefont {Zhang}(2016)}]{zhang2016berry}%
  \BibitemOpen
  \bibfield  {author} {\bibinfo {author} {\bibfnamefont {L.}~\bibnamefont
  {Zhang}},\ }\bibfield  {title} {\bibinfo {title} {Berry curvature and various
  thermal {H}all effects},\ }\href
  {https://doi.org/10.1088/1367-2630/18/10/103039} {\bibfield  {journal}
  {\bibinfo  {journal} {New Journal of Physics}\ }\textbf {\bibinfo {volume}
  {18}},\ \bibinfo {pages} {103039} (\bibinfo {year} {2016})}\BibitemShut
  {NoStop}%
\bibitem [{\citenamefont {Zhang}\ \emph {et~al.}(2019)\citenamefont {Zhang},
  \citenamefont {Zhang}, \citenamefont {Okamoto},\ and\ \citenamefont
  {Xiao}}]{zhang2019thermal}%
  \BibitemOpen
  \bibfield  {author} {\bibinfo {author} {\bibfnamefont {X.}~\bibnamefont
  {Zhang}}, \bibinfo {author} {\bibfnamefont {Y.}~\bibnamefont {Zhang}},
  \bibinfo {author} {\bibfnamefont {S.}~\bibnamefont {Okamoto}},\ and\ \bibinfo
  {author} {\bibfnamefont {D.}~\bibnamefont {Xiao}},\ }\bibfield  {title}
  {\bibinfo {title} {Thermal {H}all effect induced by magnon-phonon
  interactions},\ }\href {https://doi.org/10.1103/PhysRevLett.123.167202}
  {\bibfield  {journal} {\bibinfo  {journal} {Phys. Rev. Lett.}\ }\textbf
  {\bibinfo {volume} {123}},\ \bibinfo {pages} {167202} (\bibinfo {year}
  {2019})}\BibitemShut {NoStop}%
\bibitem [{\citenamefont {Sharma}\ \emph {et~al.}(2024)\citenamefont {Sharma},
  \citenamefont {Valldor},\ and\ \citenamefont {Lorenz}}]{sharma2024phonon}%
  \BibitemOpen
  \bibfield  {author} {\bibinfo {author} {\bibfnamefont {R.}~\bibnamefont
  {Sharma}}, \bibinfo {author} {\bibfnamefont {M.}~\bibnamefont {Valldor}},\
  and\ \bibinfo {author} {\bibfnamefont {T.}~\bibnamefont {Lorenz}},\
  }\bibfield  {title} {\bibinfo {title} {Phonon thermal {H}all effect in
  nonmagnetic {${\mathrm{Y}}_{2}{\mathrm{Ti}}_{2}{\mathrm{O}}_{7}$}},\ }\href
  {https://doi.org/10.1103/PhysRevB.110.L100301} {\bibfield  {journal}
  {\bibinfo  {journal} {Phys. Rev. B}\ }\textbf {\bibinfo {volume} {110}},\
  \bibinfo {pages} {L100301} (\bibinfo {year} {2024})}\BibitemShut {NoStop}%
\bibitem [{\citenamefont {Garate}(2013)}]{garate2013phonon}%
  \BibitemOpen
  \bibfield  {author} {\bibinfo {author} {\bibfnamefont {I.}~\bibnamefont
  {Garate}},\ }\bibfield  {title} {\bibinfo {title} {{Phonon-Induced
  Topological Transitions and Crossovers in Dirac Materials}},\ }\href
  {https://doi.org/10.1103/PhysRevLett.110.046402} {\bibfield  {journal}
  {\bibinfo  {journal} {Phys. Rev. Lett.}\ }\textbf {\bibinfo {volume} {110}},\
  \bibinfo {pages} {046402} (\bibinfo {year} {2013})}\BibitemShut {NoStop}%
\bibitem [{rqm()}]{rqmp}%
  \BibitemOpen
  \href@noop {} {}\bibinfo {howpublished}
  {\url{https://doi.org/10.69777/309032}}\BibitemShut {NoStop}%
\bibitem [{\citenamefont {Lehman}\ \emph {et~al.}(1962)\citenamefont {Lehman},
  \citenamefont {Wolfram},\ and\ \citenamefont {De~Wames}}]{lehman1962axially}%
  \BibitemOpen
  \bibfield  {author} {\bibinfo {author} {\bibfnamefont {G.~W.}\ \bibnamefont
  {Lehman}}, \bibinfo {author} {\bibfnamefont {T.}~\bibnamefont {Wolfram}},\
  and\ \bibinfo {author} {\bibfnamefont {R.~E.}\ \bibnamefont {De~Wames}},\
  }\bibfield  {title} {\bibinfo {title} {{Axially Symmetric Model for Lattice
  Dynamics of Metals with Application to Cu, Al, and Zr${\mathrm{H}}_{2}$}},\
  }\href {https://doi.org/10.1103/PhysRev.128.1593} {\bibfield  {journal}
  {\bibinfo  {journal} {Phys. Rev.}\ }\textbf {\bibinfo {volume} {128}},\
  \bibinfo {pages} {1593} (\bibinfo {year} {1962})}\BibitemShut {NoStop}%
\bibitem [{\citenamefont {Pintschovius}\ \emph {et~al.}(1983)\citenamefont
  {Pintschovius}, \citenamefont {Smith}, \citenamefont {Wakabayashi},
  \citenamefont {Reichardt}, \citenamefont {Weber}, \citenamefont {Webb},\ and\
  \citenamefont {Fisk}}]{pintschovius1983lattice}%
  \BibitemOpen
  \bibfield  {author} {\bibinfo {author} {\bibfnamefont {L.}~\bibnamefont
  {Pintschovius}}, \bibinfo {author} {\bibfnamefont {H.~G.}\ \bibnamefont
  {Smith}}, \bibinfo {author} {\bibfnamefont {N.}~\bibnamefont {Wakabayashi}},
  \bibinfo {author} {\bibfnamefont {W.}~\bibnamefont {Reichardt}}, \bibinfo
  {author} {\bibfnamefont {W.}~\bibnamefont {Weber}}, \bibinfo {author}
  {\bibfnamefont {G.~W.}\ \bibnamefont {Webb}},\ and\ \bibinfo {author}
  {\bibfnamefont {Z.}~\bibnamefont {Fisk}},\ }\bibfield  {title} {\bibinfo
  {title} {{Lattice dynamics of the $A15$ compound ${\mathrm{Nb}}_{3}$Sb}},\
  }\href {https://doi.org/10.1103/PhysRevB.28.5866} {\bibfield  {journal}
  {\bibinfo  {journal} {Phys. Rev. B}\ }\textbf {\bibinfo {volume} {28}},\
  \bibinfo {pages} {5866} (\bibinfo {year} {1983})}\BibitemShut {NoStop}%
\bibitem [{\citenamefont {Ghosh}\ and\ \citenamefont
  {Timm}(2019)}]{ghosh2019charge}%
  \BibitemOpen
  \bibfield  {author} {\bibinfo {author} {\bibfnamefont {S.}~\bibnamefont
  {Ghosh}}\ and\ \bibinfo {author} {\bibfnamefont {C.}~\bibnamefont {Timm}},\
  }\bibfield  {title} {\bibinfo {title} {Charge-spin response and collective
  excitations in {W}eyl semimetals},\ }\href
  {https://doi.org/10.1103/PhysRevB.99.075104} {\bibfield  {journal} {\bibinfo
  {journal} {Phys. Rev. B}\ }\textbf {\bibinfo {volume} {99}},\ \bibinfo
  {pages} {075104} (\bibinfo {year} {2019})}\BibitemShut {NoStop}%
\bibitem [{\citenamefont {Polini}\ \emph {et~al.}(2009)\citenamefont {Polini},
  \citenamefont {MacDonald},\ and\ \citenamefont {Vignale}}]{polini2009drude}%
  \BibitemOpen
  \bibfield  {author} {\bibinfo {author} {\bibfnamefont {M.}~\bibnamefont
  {Polini}}, \bibinfo {author} {\bibfnamefont {A.~H.}\ \bibnamefont
  {MacDonald}},\ and\ \bibinfo {author} {\bibfnamefont {G.}~\bibnamefont
  {Vignale}},\ }\href {https://arxiv.org/abs/0901.4528} {\bibinfo {title}
  {Drude weight, plasmon dispersion, and pseudospin response in doped graphene
  sheets}} (\bibinfo {year} {2009}),\ \Eprint {https://arxiv.org/abs/0901.4528}
  {arXiv:0901.4528 [cond-mat.str-el]} \BibitemShut {NoStop}%
\bibitem [{\citenamefont {Sabio}\ \emph {et~al.}(2008)\citenamefont {Sabio},
  \citenamefont {Nilsson},\ and\ \citenamefont {Castro~Neto}}]{sabio2008fsum}%
  \BibitemOpen
  \bibfield  {author} {\bibinfo {author} {\bibfnamefont {J.}~\bibnamefont
  {Sabio}}, \bibinfo {author} {\bibfnamefont {J.}~\bibnamefont {Nilsson}},\
  and\ \bibinfo {author} {\bibfnamefont {A.~H.}\ \bibnamefont {Castro~Neto}},\
  }\bibfield  {title} {\bibinfo {title} {$f$-sum rule and unconventional
  spectral weight transfer in graphene},\ }\href
  {https://doi.org/10.1103/PhysRevB.78.075410} {\bibfield  {journal} {\bibinfo
  {journal} {Phys. Rev. B}\ }\textbf {\bibinfo {volume} {78}},\ \bibinfo
  {pages} {075410} (\bibinfo {year} {2008})}\BibitemShut {NoStop}%
\bibitem [{\citenamefont {Mead}(1992)}]{mead1992geometric}%
  \BibitemOpen
  \bibfield  {author} {\bibinfo {author} {\bibfnamefont {C.~A.}\ \bibnamefont
  {Mead}},\ }\bibfield  {title} {\bibinfo {title} {The geometric phase in
  molecular systems},\ }\href {https://doi.org/10.1103/RevModPhys.64.51}
  {\bibfield  {journal} {\bibinfo  {journal} {Rev. Mod. Phys.}\ }\textbf
  {\bibinfo {volume} {64}},\ \bibinfo {pages} {51} (\bibinfo {year}
  {1992})}\BibitemShut {NoStop}%
\bibitem [{\citenamefont {Lin}\ \emph {et~al.}(2024)\citenamefont {Lin},
  \citenamefont {Liu},\ and\ \citenamefont {Lu}}]{lin2024quantum}%
  \BibitemOpen
  \bibfield  {author} {\bibinfo {author} {\bibfnamefont {H.-J.}\ \bibnamefont
  {Lin}}, \bibinfo {author} {\bibfnamefont {T.}~\bibnamefont {Liu}},\ and\
  \bibinfo {author} {\bibfnamefont {H.-Z.}\ \bibnamefont {Lu}},\ }\bibfield
  {title} {\bibinfo {title} {Quantum oscillations in acoustic phonons of
  nodal-line semimetals},\ }\href {https://doi.org/10.1103/PhysRevB.109.195421}
  {\bibfield  {journal} {\bibinfo  {journal} {Phys. Rev. B}\ }\textbf {\bibinfo
  {volume} {109}},\ \bibinfo {pages} {195421} (\bibinfo {year}
  {2024})}\BibitemShut {NoStop}%
\bibitem [{\citenamefont {Varshney}\ \emph {et~al.}(2023)\citenamefont
  {Varshney}, \citenamefont {Mukherjee}, \citenamefont {Kundu},\ and\
  \citenamefont {Agarwal}}]{varshney2023intrinsic}%
  \BibitemOpen
  \bibfield  {author} {\bibinfo {author} {\bibfnamefont {H.}~\bibnamefont
  {Varshney}}, \bibinfo {author} {\bibfnamefont {R.}~\bibnamefont {Mukherjee}},
  \bibinfo {author} {\bibfnamefont {A.}~\bibnamefont {Kundu}},\ and\ \bibinfo
  {author} {\bibfnamefont {A.}~\bibnamefont {Agarwal}},\ }\bibfield  {title}
  {\bibinfo {title} {Intrinsic nonlinear thermal {H}all transport of magnons: A
  quantum kinetic theory approach},\ }\href
  {https://doi.org/10.1103/PhysRevB.108.165412} {\bibfield  {journal} {\bibinfo
   {journal} {Phys. Rev. B}\ }\textbf {\bibinfo {volume} {108}},\ \bibinfo
  {pages} {165412} (\bibinfo {year} {2023})}\BibitemShut {NoStop}%
\end{thebibliography}%

\end{document}